\documentclass[aps,pre,preprint,onecolumn,citeautoscript,superscriptaddress,nofootinbib,eqsecnum]{revtex4-1}  
\usepackage{amsmath,amssymb,bm} 
\usepackage{graphicx}  
\usepackage[tight]{subfigure}    
\usepackage{color} 
\usepackage[papersize={8.5in,11in}]{geometry}
\usepackage[colorlinks=true]{hyperref}  
\hypersetup{
    bookmarks=true,         
    unicode=false,          
    pdftoolbar=true,        
    pdfmenubar=true,        
    pdffitwindow=false,     
    pdfstartview={FitH},    
    pdftitle={Supersymmetric SYK models},    
    pdfauthor={Wenbo Fu, Davide Gaiotto, Juan Maldacena and Subir Sachdev},     
    pdfsubject={},   
    pdfcreator={Creator},   
    pdfproducer={Producer}, 
    pdfkeywords={} {} {}, 
    pdfnewwindow=true,      
    colorlinks=true,       
    linkcolor=magenta, 
    citecolor=blue,        
    filecolor=magenta,      
    urlcolor=blue           
} 

\usepackage{amsfonts}
\usepackage{upgreek}
\usepackage{slashed}
\usepackage{latexsym}
\usepackage{braket}

\usepackage{todonotes}

\newcommand{\beq}{\begin{equation}}
\newcommand{\eeq}{\end{equation}}
\def\bea{\begin{eqnarray}}
\def\eea{\end{eqnarray}}

\newcommand{\nn}{\nonumber \\}

 \def\nref#1{(\ref{#1})}
 \def\be{\begin{equation}}
 \def\ee{\end{equation}}
 \def\sign{{\rm sgn}}
\def\half{{1 \over 2 }}
 
\def\la{\label}

\begin{document}

\title{Supersymmetric SYK models}

\author{Wenbo Fu}

\affiliation{Department of Physics, Harvard University, Cambridge MA 02138, USA}

\author{Davide Gaiotto}

\affiliation{Perimeter Institute for Theoretical Physics, Waterloo, Ontario, Canada N2L 2Y5}

\author{Juan Maldacena}

\affiliation{Institute for Advanced Study, Princeton, NJ 08540, USA}

\author{Subir Sachdev}

\affiliation{Department of Physics, Harvard University, Cambridge MA 02138, USA}
\affiliation{Perimeter Institute for Theoretical Physics, Waterloo, Ontario, Canada N2L 2Y5}

\date{\today
\\
\vspace{0.4in}}

\begin{abstract}%

We discuss a supersymmetric generalization of the Sachdev-Ye-Kitaev model. These are quantum mechanical models 
involving $N$ Majorana fermions. The supercharge is given by a polynomial expression in terms of the Majorana fermions with 
random coefficients. The Hamiltonian is the square of the supercharge. 
The ${\cal N}=1$ model with a single supercharge has unbroken supersymmetry at large $N$, but non-perturbatively 
spontaneously broken supersymmetry 
in the exact theory. We analyze the model by looking at the large $N$ equation, and also by performing numerical computations for small values of $N$. 
We also compute the large $N$ spectrum of ``singlet'' operators, where we find a structure qualitatively similar to the ordinary SYK model. 
 We also discuss an ${\cal N}=2$ version. In this case, the model preserves supersymmetry in the exact theory and we can compute a 
 suitably weighted Witten index to count the number of ground states, which agrees with the large $N$ computation of  the entropy. 
 In both cases, we discuss the supersymmetric generalizations of the Schwarzian action which give the dominant effects at low energies. 
 
\end{abstract}

\maketitle

\section{Introduction}
\label{sec:intro}

The Sachdev-Ye-Kitaev (SYK) models (or their variants) realize non-Fermi liquid states of matter without quasiparticle excitations \cite{SY92,PG98,GPS99,GPS01}.
They also have features in common with 
 black holes with AdS$_2$ horizons \cite{SS10,SS10b}, and this connection has been significantly
sharpened in recent work \cite{kitaev2015talk,AAJP15,AABK16,SS15,Hosur15,JPRV16,YLX16,WFSS16,Jevicki16,Jevicki16b,JMDS16,JMDS16b,DHT16,Garcia-Alvarez:2016wem,HV16,KJ16,BAK16,CP16,GQS16}. 

In this paper, we introduce supersymmetric generalizations of the SYK models. Like previous models, the supersymmetric models
have random all-to-all interactions between fermions on $N$ sites. There are no canonical bosons in the underlying Hamiltonian,
and in this respect, our models are similar to the supersymmetric 
lattice models in Refs.~\cite{FSB03,FNS03,FS05,Liza08,Liza09,Liza11,Liza12}.
As we describe below, certain structures in the
correlations of the random couplings of our models lead to $\mathcal{N}=1$ and $\mathcal{N}=2$ supersymmetry. 
Supersymmetric models with random couplings that include both bosons and fermions were considered in \cite{Anninos:2016szt}.

Let us  discuss now  the model with $\mathcal{N}=1$ supersymmetry, and defer presentation of the $\mathcal{N}=2$ case
to Section~\ref{sec:N2}. For the $\mathcal{N}=1$ case, we introduce the supercharge
\beq
Q=i \sum_{1\leqslant i<j<k\leqslant N} C_{ijk}\psi^i\psi^j\psi^k \, ,
\label{defQ}
\eeq
where $\psi_i$ are Majorana fermions on sites $i=1 \ldots N$, 
\beq
\{\psi^i , \psi^j \} = \delta^{ij},
\eeq
and $C_{ijk}$ is a fixed real  $N\times N\times N$ antisymmetric tensor so that $Q$ is Hermitian. 
We will take the $C_{ijk}$ to be independent gaussian random variables, with zero mean and variance specified by the constant $J$: 
\beq
\overline{C_{ijk}}=0,\quad ~~~~~~~ \overline{C_{ijk}^2}= { 2 J \over N^2 }  
\label{CCaverage}
\eeq
where $J$ is positive and has units of energy. 
As is the case in supersymmetric theories, the Hamiltonian is the square of the supercharge
\beq
\mathcal{H}=Q^2=
 E_0 +\sum_{1\leqslant i<j<k<l\leqslant N}J_{ijkl} \, \psi^i\psi^j\psi^k\psi^l
\label{SuperSYKH}
\eeq
where
\beq
 E_0=\sum_{1\leqslant i<j<k\leqslant N}C_{ijk}^2 \quad , \quad  J_{ijkl}=-\frac{1}{8}\sum_a C_{a[ij}C_{kl]a},
\eeq
with $[\,]$ representing all possible anti-symmetric permutations. Note that the $J_{ijkl}$ are not independent gaussian random variables,
and this is formally the only difference from the Hamiltonian of the non-supersymmetric SYK models
\cite{kitaev2015talk,SS15,Hosur15,JPRV16,YLX16,WFSS16,Jevicki16,Jevicki16b,JMDS16,DHT16,BAK16}. 
These particular correlations change the structure of the large $N$ equations and lead to a solution where the fermion has dimension 
$\Delta_f = 1/6 $. In addition, there is a supersymmetric partner of this operator which is bosonic and has dimension $\Delta_b = 2/3 = 1/2 + \Delta_f$. 
This large $N$ solution has unbroken supersymmetry, and we have checked this numerically by comparing with exact diagonalization of the Hamiltonain. 
We have also computed the large $N$ ground state entropy from a complete numerical solution of the saddle-point equations. In the exact diagonalization we find that the 
lowest energy state has non-zero energy, and therefore, broken supersymmetry. However, this energy is estimated to be of order $e^{ - \alpha N }$ where $\alpha $ is a numerical
constant. 
We have also generalized the model to include a supercharge of the schematic form $Q \sim \psi^{\hat q }$, and we also solved this model in the large $\hat q$ limit. 
We also formulated the model in superspace, and show that the large $N$ equations have a super-reparameterization invariance, which is both 
spontaneously as well as explicitly broken by the appearance of a superschwarzian action, which we describe in   detail. 

 We have also analyzed the  eigenvalues of the ladder kernel which appears in the computation of the four point function. There are both bosonic and fermionic
 operators that can propagate on this ladder. There is a particular eigenvalue of the kernel which is a zero mode and corresponds to the degrees of freedom 
 described by the Schwarzian. They are a bosonic mode with dimension $h=2$ and a fermionic one with $h=3/2$. The other eigenvalues of the kernel should 
 describe operators appearing in the OPE. These also come in boson-fermion pairs and have a structure similar to the usual SYK case. One interesting feature is
 the appearance of a boson fermion pair with dimensions $h=1$ and $h= {3 \over 2 }$, which is associated to an additional 
  symmetry of the low energy equations. These do not give rise to extra zero modes but simply correspond  to other operators in the theory.  
 
   We have also analyzed the ${\cal N}=2$ version of the theory. In this case we can also compute a kind of Witten index. More precisely, the model has  a 
 discrete $Z_{\hat q }$ global symmetry that commutes with supersymmetry, so that we can include the corresponding discrete chemical potential in the Witten index, 
 which turns out to be non-zero. These are generically expected to be lower bounds on the large $N$ ground state entropy; it turns out that the largest Witten index is, 
 in fact, equal to the large $N$ ground state entropy. The model also has a $U(1)_R$ symmetry. 
 The exact diagonalization analysis also suggests a conjecture for number of ground states for each value of the   $U(1)_R$ charge. For the $\hat q=3$ case, they
 are concentrated at very small values of the $U(1)$ R-charge, within $|Q|\leq 1/3 $.

   This paper is organized as follows. In section II we define the ${\cal N}=1$ supersymmetric model, write the large $N$ effective action and  the corresponding 
   classical equations.  We determine the dimensions of the operators in the IR and we derive a  constraints imposed by unbroken supersymmetry on the 
   correlators. We also present a generalization of the model where the supercharge is a product of $\hat q$ fermions and solve the whole flow in the $\hat q \to \infty $ 
   limit. In section III we present some results on exact numerical diagonalization of the Hamiltonian. This includes results on the ground state energy and two point correlation functions. In section IV we discuss the physics of the low energy  degrees of freedom associated to the spontaneously and explicitly broken super-reparameterization symmetry of the theory. In section V we define and study a model with ${\cal N}=2$ supersymmetry. We compute the Witten index and use it to argue that the model has a large exact degeneracy at zero energy. We also discuss the superspace and super-reparameterization symmetry in this case. 
    In section VI we discuss the ladder  diagrams that contribute to the four point function. We use them to determine the eigenvalues of the ladder kernel and use it
    to determine the spectrum of dimensions of composite  operators. 
    
\section{Definition of the model and the large $N$ effective action}
\label{sec:eff}

To set up a path integral formulation of $H$, we first note that the supercharge acts on the fermion as
\beq
\{Q,\psi^i\}=i \sum_{1\leqslant j<k\leqslant N} C_{ijk}\psi^j\psi^k .
\eeq
We introduce a non-dynamical auxiliary boson $b^i$ to linearize the supersymmetry transformation and realize the supersymmetry algebra off-shell. The Lagrangian
describing $H$ is 
\beq
\mathcal{L}=\sum_i \left[ \frac{1}{2}\psi^i\partial_{\tau}\psi^i-\frac{1}{2}  b^ib^i+ i  \sum_{1\leqslant j<k\leqslant N}C_{ijk}b^i\psi^j\psi^k \right] .\label{Leff}
\eeq
Under the transformation $Q \psi^i = b^i$, $Q  b^i =  \partial_\tau \psi^i$ it changes as 
\beq
Q \mathcal{L}=\partial_{\tau}\left( - \frac{1}{2}  \sum_i\psi^i b^i + \frac{i}{3} \sum_{1\leqslant j<k\leqslant N}C_{ijk}\psi^i\psi^j\psi^k\right) + i \sum_{1\leqslant j<k\leqslant N}C_{ijk}b^i b^j\psi^k .\label{Leffch}
\eeq
This implies that the action is invariant as long as the structure constants $C_{ijk}$ in \nref{Leff} are totally anti-symmetric. 

Now we proceed to obtain the effective action. This can be done by averaging over the 
Gaussian random variables $C_{ijk}$ in the replica formalism. In this model, as in SYK, the interaction between replicas is suppressed by $1/N^2$, so that 
we can simply average over disorder by treating it as an additional field with time indepedent two point functions as in \nref{CCaverage}. 
Averaging over disorder, we obtain
\begin{align} \la{FerAct}
S_{\rm eff}&=\int_0^\beta d\tau (\frac{1}{2}\psi ^i\partial_\tau\psi ^i-\frac{1}{2}b^i b^i )-\frac{J }{N^2}\int_0^\beta d\tau_1 d\tau_2 \big(b^i (\tau_1)b^i (\tau_2)\big)\big(\psi^j (\tau_1)\psi^j (\tau_2)\big)^2 +\cr
&-\frac{2J }{N^2}\int_0^\beta d\tau_1 d\tau_2 \big(b^i (\tau_1)\psi^i (\tau_2)\big)\big(\psi^j (\tau_1)b^j (\tau_2)\big)\big(\psi^k (\tau_1)\psi^k (\tau_2)\big) .
\end{align}
Note that this action contains terms in which the bosons and fermions carry the same index, and which should be omitted \textit{e.g.} $b^i (\tau_1)b^i (\tau_2)\psi^i (\tau_1)\psi^i (\tau_2)\psi^j (\tau_1)\psi^j (\tau_2)$;
however they are subdominant in the large $N$ limit, and so we  ignore this issue.

Notice further that the relative coefficient between the last two terms is determined by the supersymmetry requirement that the structure constants $C_{ijk}$ are totally anti-symmetric,
so that 
\begin{equation}
\langle C_{ijk} C_{i' j' k'}  \rangle \sim \delta_{ii'} \delta_{jj'} \delta_{kk'} +  \delta_{ij'} \delta_{jk'} \delta_{ki'} + \delta_{ik'} \delta_{ji'} \delta_{kj'} + (j \leftrightarrow k)
\end{equation}

The purpose of this section is to discuss the large $N$ saddle-point equations for the diagonal   
Green's functions
\begin{align}
G_{\psi \psi}(\tau_1,\tau_2)&=\frac{1}{N}\psi^i (\tau_1)\psi^i(\tau_2),\cr 
G_{b b}(\tau_1,\tau_2)& =\frac{1}{N}b^i (\tau_1)b^i (\tau_2),
\end{align}
where we have a sum over $i$. We will thus drop the last term in (\ref{FerAct}), which only affects the saddle-point equations for the off-diagonal  
Green's functions
\begin{align}
G_{b \psi}(\tau_1,\tau_2)& =\frac{1}{N}b^i (\tau_1)\psi^i (\tau_2),\cr 
G_{\psi b}(\tau_1,\tau_2)& =\frac{1}{N}\psi^i (\tau_1)b^i (\tau_2),
\end{align}
We will restore it in a later section \ref{sec:superspace}, where we write the saddle-point equations in a manifestly super-symmetric fashion. 

We introduce the Lagrange multipliers $\Sigma_{\psi \psi}$
\beq
1 =\int \mathcal{D} G_{\psi \psi}  \mathcal{D}\Sigma_{\psi \psi} \exp\Big(-\frac{N}{2}\Sigma_{\psi \psi}(\tau_1,\tau_2)\big(G_{\psi \psi}(\tau_1,\tau_2)-\frac{1}{N}\psi^i (\tau_1)\psi^i (\tau_2)\big)\Big),
\eeq
and $\Sigma_{bb}$
\beq
1 =\int \mathcal{D} G_{bb}  \mathcal{D}\Sigma_{b b} \exp\Big(-\frac{N}{2}\Sigma_{b b}(\tau_1,\tau_2)\big(G_{b b}(\tau_1,\tau_2)-\frac{1}{N}b^i (\tau_1)b^i (\tau_2)\big)\Big).
\eeq
As the notation suggests, these Lagrange multipliers will eventually become the self energies.
Inserting these factors of 1 in the fermion path integral with the action \nref{FerAct}, using the delta functions implied by the integration over $\Sigma_{\psi,b}$ to
express the interaction terms in \nref{FerAct}, and integrating out the fermions we obtain 
\bea
&& Z = \int \mathcal{D} G_{\psi \psi}  \mathcal{D}\Sigma_{\psi \psi}  e^{ -   S_{eff} (G_{\psi},G_{bb},\Sigma_{\psi \psi},\Sigma_{bb}) } 
\cr
&& S_{\rm eff}(G_{\psi},G_{bb},\Sigma_{\psi \psi},\Sigma_{bb})/N = -\log{\text{Pf}[\partial_{\tau}-\Sigma_{\psi \psi}(\tau)]}+ \half \log{
\det [-1-\Sigma_{bb}(\tau)] } +\label{seff}\\
&~&~~~~~ + \half \int d\tau_1 d\tau_2 \left[ 
 \Sigma_{\psi \psi}(\tau_1,\tau_2)G_{\psi \psi}(\tau_1,\tau_2)+ \Sigma_{b b}(\tau_1,\tau_2)G_{bb}(\tau_1,\tau_2)
-{ J \  }  G_{bb}(\tau_1,\tau_2)G_{\psi \psi }(\tau_1,\tau_2)^2 \right] ,
 \nonumber
\eea
which becomes a classical action when $N$ is large. 
Let us look at the  classical equations for   the action in (\ref{seff}). 
Taking derivatives with respect to $G_{\psi}$ and $G_{bb}$, we obtain
\bea
\Sigma_{\psi \psi}(\tau_1,\tau_2)&=& 2J G_{bb}(\tau_1,\tau_2)G_{\psi \psi}(\tau_1,\tau_2)\nonumber\\
\Sigma_{bb}(\tau_1,\tau_2)&=&J G_{\psi \psi}(\tau_1,\tau_2)^2 ,\label{EOM12}
\eea
Taking derivatives with respect to $\Sigma_{\psi \psi}$ and $\Sigma_{bb}$, assuming time translation symmetry and going to Fourier space, 
 we obtain
\bea
G_{\psi}(i\omega)^{-1}&=&-i\omega-\Sigma_{\psi \psi}(i\omega)\nonumber\\
G_{bb}(i\omega)^{-1}&=&-1-\Sigma_{bb}(i\omega), \label{EOM34}
\eea
which confirms that $\Sigma_{\psi,b}$ are the self energies.  

In temporal space, the saddle point equations take the form 
\bea
\partial_{\tau_1} G_{\psi \psi}(\tau_1, \tau_3) - \int d\tau_2 \left( 2 J G_{bb}(\tau_1, \tau_2)G_{\psi \psi}(\tau_1, \tau_2) \right)G_{\psi \psi}(\tau_2, \tau_3)&=&\delta(\tau_1- \tau_3) \nonumber\\
-G_{bb}(\tau_1, \tau_3) - \int d\tau_2 \left(J G_{\psi \psi}(\tau_1, \tau_2)^2 \right)G_{bb}(\tau_2, \tau_3)&=&\delta(\tau_1- \tau_3), \label{EOM56}
\eea

\begin{figure}[h]
\center
\includegraphics[width=3.5in]{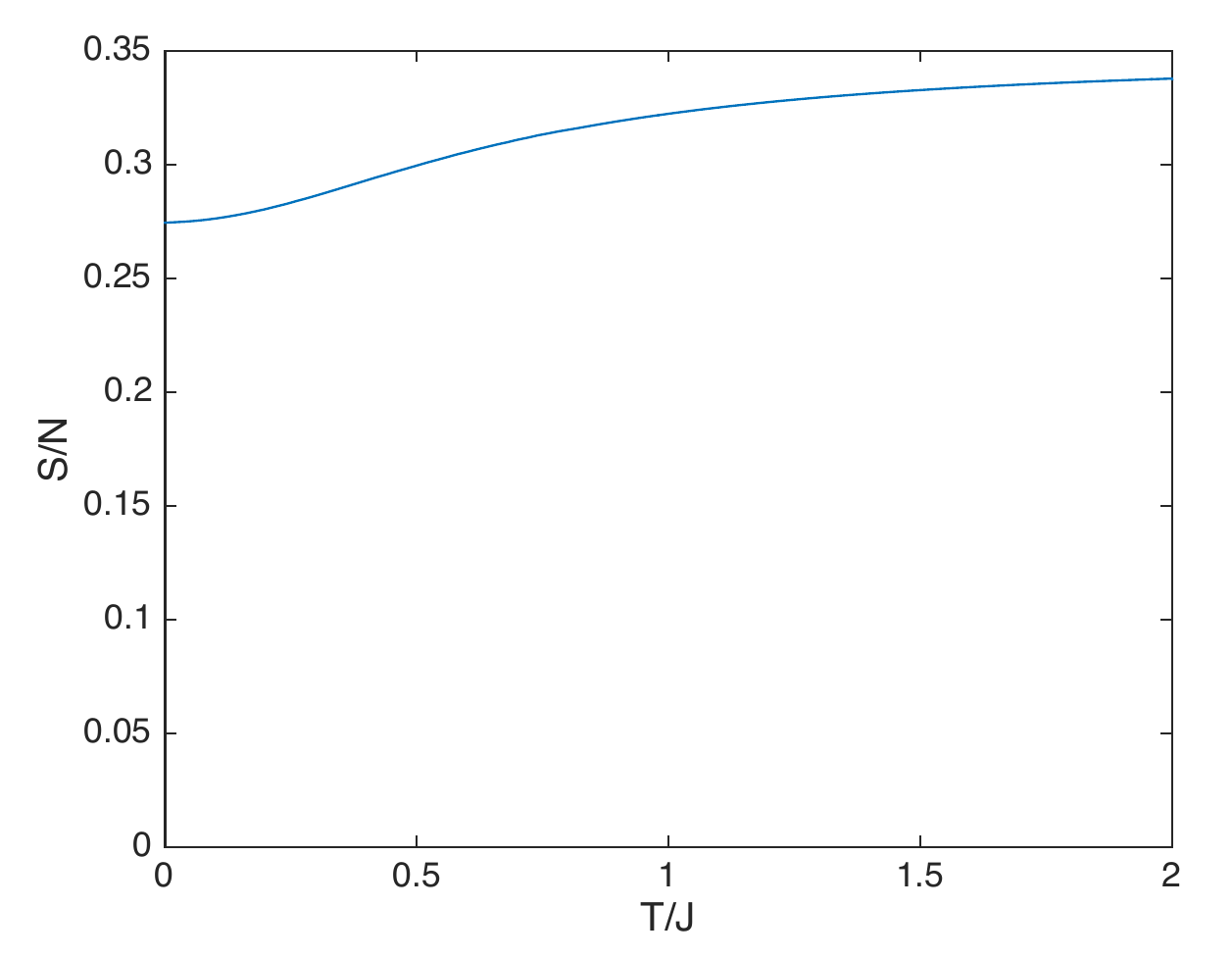}
\caption{Thermal entropy obtained by numerically  solving the large $N$ equations of motion \nref{EOM12}\nref{EOM34}.
 At high temperatures we have just the log of the dimension of the Hilbert space, 
${S \over N } = \half \log 2 $. The zero temperature entropy is approximately ${S \over N } \sim 0.2745+0.0005$, where the error is estimated by the convergence of the FFT (Fast Fourier Transform) algorithm. The analytical result $
 {S \over N } =\frac{1}{2}\log{\left[2\cos{\frac{\pi}{6}}\right]} $ \nref{GSEntropy} also lies in this range  .}
\label{Fig:Entropy}
\end{figure}

These equations can be solved numerically, and we can see some plots in figure \nref{Fig:GpsiGbfiniteT}. 
Once we find a solution to these equations, we can compute the on-shell action, which can be written as 
\bea
{ \log Z \over N } &=&  -{ S_{eff} \over N } 
=\frac{1}{2}\log{2} -
\sum_{n\in \text{half integer}}\frac{1}{2}\log{\left[-i\omega_n G_{\psi \psi}(i\omega_n)\right]} + \sum_{n\in \text{integer}} \frac{1}{2}\log{G_{bb}(i\omega_n)} 
\cr
&& ~~~~- { J \beta \over 2 }  \int_0^\beta G_{bb}(\tau)G_{\psi \psi}(\tau)^2 
\eea 
where $\omega_n$ are the Matsubara fequencies for the fermion and boson cases. 
From this we can compute the entropy through the usual thermodynamic formula. 
A plot of the entropy as a function of the temperature can be found in figure \nref{Fig:Entropy}. 

We can now  determine the low energy structure of the solutions of (\ref{EOM12}) and (\ref{EOM34}), as in 
 \cite{SY92}, by making a power law ansatz at late times  ($ 1 \ll J \tau \ll N $)
\beq
G_{\psi\psi} \propto \frac{1}{\tau^{2\Delta_{\psi}}} \quad , \quad G_{bb} \propto \frac{1}{\tau^{2\Delta_{b}}}, \label{longtime}
\eeq
where $\Delta_\psi$ and $\Delta_b$ are the scaling dimensions of the fermion and the boson.
We then insert (\ref{longtime}) into (\ref{EOM12}), (\ref{EOM34}) in order  to fix the values of $\Delta_\psi$ and $\Delta_b$.
Matching
the power-laws in the saddle point equations
  yields only  the single constraint
\beq
2\Delta_{\psi}+\Delta_b=1 .\label{scaling1}
\eeq
As we will see later the dimension can be determined by looking at the constant coefficients. Before showing this, let us discuss a simpler 
way to obtain another condition. 

\subsection{Supersymmetry constraints }

Further analytic progress can be made if we assume that the solutions of the saddle point equations (\ref{EOM12}), (\ref{EOM34})
preserve supersymmetry. With such an assumption, we now show that the scaling dimensions $\Delta_\psi$ and $\Delta_b$ 
can be easily determined. Again, we refer the reader to a later section \ref{sec:superspace}  for a full discussion of the supersymmetry properties of the saddle point equations. 

If supersymmetry is not spontaneously broken, then
\begin{eqnarray} 
G_{bb}(\tau_1-\tau_2) &=& \langle b(\tau_1)b(\tau_2)\rangle=\langle Q\psi(\tau_1) b(\tau_2)\rangle=\langle\psi(\tau_1)Qb(\tau_2)\rangle
\notag 
\\
&=&\partial_{\tau_2}\langle\psi(\tau_1)\psi(\tau_2)\rangle = -\partial_{\tau_1}G_{\psi \psi}(\tau_1-\tau_2) \label{SUSYequation}
\end{eqnarray}

This relationship together with 
\beq
\Sigma_{\psi \psi}(\tau_1-\tau_2)=-\partial_{\tau_1}\Sigma_{bb}(\tau_1-\tau_2). \label{SUSYSigma}
\eeq
is compatible with the saddle-point equations in Section~\ref{sec:eff}.

 \nref{SUSYequation}  together with the ansatz \nref{longtime} leads to  
 %
\beq
\Delta_{b}= \Delta_{\psi} + \half . \label{SusyW}
\eeq
 Together with Eq.~(\ref{scaling1}), we can now determine the scaling dimensions
 \beq
 \Delta_{\psi}=\frac{1}{6} \quad , \quad \Delta_b=\frac{2}{3}. \label{eq:dim}
 \eeq
 
 \subsection{Simple  generalization }
 
We now show how to derive the $\Delta_b = \Delta_\psi + \half$ constraint directly from the saddle point equations 
without assuming that the solution preserves supersymmetry. 

It is useful to consider a simple generalization of   Eq.~(\ref{defQ}) to case where the supercharge $Q$ is the sum over products of $\hat q$ fermions\footnote{
In detail $Q = i^{ \hat q -1 \over 2 } \sum_{j_1 < j_2 < \cdots j_{\hat q} } C_{j_1,j_2 \cdots,j_n} \psi^{i_1} \psi^{i_2} \cdots \psi^{i_{\hat q}} $, with 
$\langle C_{j_1 , j_2, \cdots, j_{\hat q} }^2 \rangle = { (\hat q -1)! J \over N^{\hat q -1} }$.}. The Hamiltonian $\mathcal{H}=Q^2$ involves sums of terms with 
up to $ 2\hat q -2$ fermions. $\hat q=3$ corresponds to the case discussed above \nref{defQ}. 
 
 The large $N$  equations are \nref{EOM34} and
 \be   \label{gqeq}
 \Sigma_{\psi \psi}(\tau_1,\tau_2)=(\hat q -1)  J G_{bb}(\tau_1,\tau_2)G_{\psi}(\tau_1,\tau_2)^{\hat q -2}~,~~~~~~~~~
\Sigma_{bb}(\tau_1,\tau_2)=   J G_{\psi \psi}(\tau_1,\tau_2)^{\hat q -1} 
\ee
 We can explore them at low energy by 
making  the ansatz 
\be \la{ansatzq}
 G_{\psi \psi}(\tau_1,\tau_2) = { b_\psi \sign(\tau_{12} ) \over |\tau_{12}|^{ 2 \Delta_\psi } }~,~~~~G_{bb}(\tau_1,\tau_2) = { b_b \over |\tau_{12}|^{2 \Delta_b } }~,~~~~~~~~~~~
\tau_{12} \equiv  \tau_1 -\tau_2 
\ee
where $b_\psi$, $b_b$ are some constants. 

Again, if we assume supersymmetry we immediately derive $\Delta_\psi = 1/(2 \hat q )$ and $\Delta_b = \Delta_\psi + {1 \over 2 } $. 
Doing so without that assumption requires us to look at the equations for $b_\psi$ and $b_b$.

Using the Fourier transforms for symmetric and antisymmetric functions 
\be
\int dt e^{ i \omega t } { \sign(t) \over |t|^{ 2 \Delta } }  = c_f(\Delta) \sign(\omega ) |\omega|^{ 2 \Delta - 1 } ~,~~~~~~~~
\int dt e^{ i \omega t } { 1 \over |t|^{ 2 \Delta } }  = c_b(\Delta)   |\omega|^{ 2 \Delta - 1 } ~,~~~~~~~~
\ee
\be
c_f(\Delta ) \equiv 2 i   \cos (\pi \Delta) \Gamma(1-2 \Delta)  ~,~~~~~~~~~~c_b(\Delta) \equiv  2 \sin (\pi \Delta) \Gamma(1-2 \Delta)
\ee
The following relations are useful
\be
c_f(\Delta ) c_f(1-\Delta ) \equiv - \frac{2 \pi \cos \pi \Delta}{(1- 2 \Delta) \sin \pi \Delta}  ~,~~~~~~~~~~c_b(\Delta)c_b(1-\Delta) \equiv  - \frac{2 \pi \sin \pi \Delta}{(1- 2 \Delta) \cos \pi \Delta}
\ee

Then \nref{gqeq} , together with the low energy aproximation of \nref{EOM34}, which is $G_{\psi}(i\omega) \Sigma_{\psi \psi}(i\omega) = -1$ and
, $G_{b}(i\omega) \Sigma_{b b}(i\omega) = -1$,
gives the conditions 
\bea  
&&   J b_\psi^{\hat q -1} b_b (\hat q -1) c_f(\Delta_\psi) c_f((\hat q -2)\Delta_\psi + \Delta_b)  |\omega|^{2 (\hat q -1)\Delta_\psi + 2\Delta_b-2} =-1   ~,~~~~~~~ 
\cr
& & J b_\psi^{\hat q -1} b_b  c_b(\Delta_b) c_b((\hat q -1) \Delta_\psi)   |\omega|^{2(\hat q -1) \Delta_\psi + 2 \Delta_b  -2} =-1 \la{LEEq}
 \eea
Matching the frequency dependent part we get the condition $\Delta_b = 1 - ( \hat q -1) \Delta_\psi$. The equations for the coefficients reduce to 
 \bea  
&&  2 \pi  J b_\psi^{\hat q -1} b_b (\hat q -1) =(1- 2 \Delta_\psi)  \frac{\sin \pi \Delta_\psi}{\cos \pi \Delta_\psi}   ~,~~~~~~~ 
\cr
& & 2 \pi J b_\psi^{\hat q -1} b_b =(2 ( \hat q -1) \Delta_\psi -1)  \frac{\sin \pi  ( \hat q -1)\Delta_\psi} {\cos \pi  ( \hat q -1)\Delta_\psi}  \la{LEEq2}
 \eea
 
The ratio between the two equations gives another condition for $\Delta_\psi$, with one rational
solution obeying $\Delta_b = \Delta_\psi + \half $, which is also independently implied by supersymmetry, see  \nref{SUSYSigma}. 
In the range where $\Delta_\psi$ and $\Delta_b$ are both positive there is a second, irrational solution to the equations which has 
higher $\Delta_\psi$. This second solution breaks supersymmetry, since it does not obey \nref{SusyW}. It would be nice to understand it further, but we leave that to the future. 

We also see that the low energy  equations have a symmetry 
\be \la{symNai} 
G_{\psi \psi} \to \lambda^2  G_{\psi \psi} ~,~~~~~~~~~G_{bb} \to \lambda^{ 2 - 2 \hat q }  G_{bb}
\ee
Indeed \nref{LEEq} involves  only the combination $Jb_\psi^{\hat q -1} b_b$. This symmetry of the IR equations is 
broken by the UV boundary conditions that arise from considering the full equations in \nref{EOM34}. 

In fact, the supersymmetry relation \nref{SUSYSigma} also 
fixes this freedom of rescaling, by setting $b_b = 2 \Delta_\psi b_\psi$. 
  
In the end this fixes the coefficients to 
\be \la{Jbb}
J b_\psi^{ \hat q -1} b_b = { \tan{ \pi \over 2 \hat q } \over 2 \hat q \pi } ~,~~~~~b_b= { 1 \over \hat q } b_\psi ~,~~~~\Rightarrow ~~ 
b_\psi = \left[ {\tan{ \pi \over 2 \hat q } \over 2 \pi J }\right]^{ 1 \over \hat q } 
 \ee
 This coefficient (for $\hat q=3$) is used in the plot of figure \ref{fig:GGc}.
Of course the finite temperature version is 
\be \la{FinT} 
G_{\psi \psi} (\tau) = b_\psi \left[ \pi \over \beta \sin{ \pi \tau \over \beta } \right]^{2 \Delta_\psi}  
\ee
This generalization makes it easy to compute the ground state entropy. In principle this can be done by inserting these solutions into the 
effective action 
\be
{ \log Z \over N } = { 1 \over 2 } \log \det ( \partial_\tau - \Sigma_{\psi \psi}) - { 1 \over 2 } \log \det  ( \delta - \Sigma_{bb} ) 
+ { 1 \over 2 } \int d\tau d\tau' \left[ - \Sigma_{bb} G_{bb} - \Sigma_{\psi \psi} G_{\psi \psi}+ { J    }  G_{bb} G_{\psi \psi}^{\hat q -1 }  \right]
\ee
It is slighly simpler to take the derivative with respect to $\hat q$, ignoring any term that involves derivatives of $G_{b,\psi}$ since those terms vanish
by the equations of motion. This gives 
\be \la{derfree}
\partial_{\hat q } { \log Z \over N } =   {   J  \over 2 } \beta \int d\tau G_{bb}(\tau) G_{\psi \psi}^{\hat q -1 } \log G_{\psi \psi} =  
 \beta ( {\rm constant} )  +  {  \pi^2  \over 2 \hat q } {   J b_b b_\psi^{ \hat q -1}    }  = \beta ( { \rm constant} ) + { \pi \tan { \pi \over 2 \hat q } \over 4 \hat q^2 } 
\ee
Where we inserted  \nref{FinT} and used \nref{Jbb}. 
The constant term includes UV divergencies which are $\beta$ independent. This term contributes   to the ground state energy\footnote{If we computed it using the 
exact solution (as opposed to the conformal soution) of the equations we expect the ground state energy  to vanish due to supersymmetry.},  but
  not   to the
entropy.  
Integrating \nref{derfree} we obtain the ground state entropy 
\be \la{GSEntropy}
{ S \over N } = \half \log[ 2 \cos \pi \Delta_f] = \half \log[ 2 \cos { \pi \over 2 \hat q } ] 
\ee
where in integrating we used the boundary condition that the entropy should be the entropy of free fermion system at $\hat q =\infty$, a fact we will check below. 
For $\hat q=3$ this matches the numerical answer, see figure \ref{Fig:Entropy}. 


\subsection{The large $\hat q $ limit }

It is interesting to take the large $\hat q$ limit of the model since then we can find an exact solution 
interpolating between the short and long distance behavior. The analysis is very similar to the one in \cite{JMDS16}. 
We expand the functions as follows 
\be
G_{\psi \psi}(\tau) = { 1\over 2 } \epsilon(\tau)  +  { 1 \over 2 \hat q } g_{\psi \psi}(\tau)  ~,~~~~~~~~~G_{bb}(\tau) = - \delta(\tau) + {1\over 2 \hat q } g_{bb} 
 \label{Gqexp} \ee
 where we neglected higher order terms in the $1/\hat q $ expansion. 
 We can then Fourier transform, compute $\Sigma_{\psi \psi}$, $\Sigma_{bb}$ to first order in the ${1 \over \hat q}$ expansion. This gives $\Sigma_{\psi \psi}(i\omega) = 
 { \omega^2 \over 2 q} [ \sign g_{\psi \psi}](i\omega) $, and $\Sigma_{bb}(i \omega) ={ 1 \over 2 \hat q } g_{bb}(i \omega)$. 
  Replacing this   the equations \nref{gqeq} we find 
  \be
  \partial_\tau^2 g_{\psi \psi} =  {\cal J }^2 e^{ 2 g_{\psi \psi} } ~,~~~~~g_{bb} = { \cal J } e^{g_{\psi \psi} } ~,~~~~~~~~~{ \cal J } \equiv { \hat q J \over 2^{ \hat q - 2 } } 
  \ee
  where we take the large $\hat q$ limit keeping ${\cal J }$ fixed. 
  The solution obeying the boundary conditions $g_{\psi \psi}(0) = g_{\psi \psi}(\beta ) =0$ is 
  \be \la{soleq}
  e^{ g_{\psi \psi}} = { 1 \over \beta {\cal J } } { v \over \sin ( v { \tau \over \beta }  + b)}   ~,~~~~~{ \beta \cal J } = { v \over \cos { v \over 2 } }~,~~~~~~b = { \pi - v \over 2} 
  \ee
 where $v,~b$ are integration constants fixed by the boundary  conditions.  
  It is interesting to note that the UV supersymmetry condition $g_{bb} = - \partial_\tau g_{\psi \psi}$ is only approximately true 
  at short distances, distances shorter than the temperature. 
  
  It is also interesting to compute the free energy. Again, this is conveniently done by taking a derivative with respect to $J$ and using the equations of motion. 
  \be \la{DerF}
  {  \cal  J } \partial_{ \cal  J } { \log Z \over N } = - \left. { \beta \over 2 ( \hat q -1)} \partial_\tau G_{\psi \psi} \right|_{\tau = 0^+}=
   - { \beta \over 2 \hat q^2 } \partial_\tau g_{\psi \psi }   |_{\tau = 0^+}
   \ee
   where the first equality holds in general and the second only for large $\hat q$. Expressing it in term of the parameters in \nref{soleq} we get
   \bea \la{FreeLq}
   { \log Z \over N } &= & { 1 \over 2 } \log 2 + { 1 \over 4 \hat q^2 } \left( - { v^2 \over 4 } + v \tan{ v \over 2 } \right) ~,~~~~~{\rm with}~~~~~{ \beta \cal J } = { v \over \cos { v \over 2 } }
   \cr
   &\sim & { 1 \over 2 } \log 2 + { 1 \over   \hat q^2 } \left[ { \beta {\cal J } \over 4 } - { \pi^2 \over 16 } + { \pi^2 \over 8 \beta {\cal J } }  - { \pi^2 \over 4 (\beta {\cal J})^2 } + \cdots \right] ~,~~~~~~{\rm for}
   ~~~ {\beta {\cal J } } \gg 1 
 \eea
 we can also easily compute the small $(\beta {\cal J})$ expansion, which, as expected, goes in powers of $(\beta {\cal J })^2 $. 
  We have used the entropy of the free fermion system,  at $\beta {\cal J } \to 0$, as an integration constant in going from \nref{DerF} to \nref{FreeLq}. 
  The constant term in the large $(\beta {\cal J })$ expansion agrees with the large $\hat q$ expansion of the ground state entropy \nref{GSEntropy}. 
  The $1/(\beta {\cal J } )$ term can also be obtained form the Schwarzian and this can serve as a way to fix the coeffiicient of the Schwarzian action at large $\hat q$. 
  The linear term in ${\beta {\cal J } }$ represents the ground state energy and it should be subtracted off. 
  
  All these results have the same form as the large $q$ limit of the usual SYK model \cite{JMDS16}. This is not a coincidence. What happens is that the leading boson 
  propagator is simply the delta function in \nref{Gqexp} which collapses the diagrams to those of the large $q$ limit of the usual SYK model.

\section{Exact diagonalization}

This section presents results from the exact numerical diagonalization of the Hamiltonian in Eq.~(\ref{SuperSYKH}).
We examined samples with up to $N=28$ sites, and averaged over 100 or more realizations of disorder.
This exact diagonalization allows us to check the validity of the answer we obtained using large $N$ methods.
 
\subsection{Supersymmetry}

\begin{figure}[h]
\center
\includegraphics[width=3.5in]{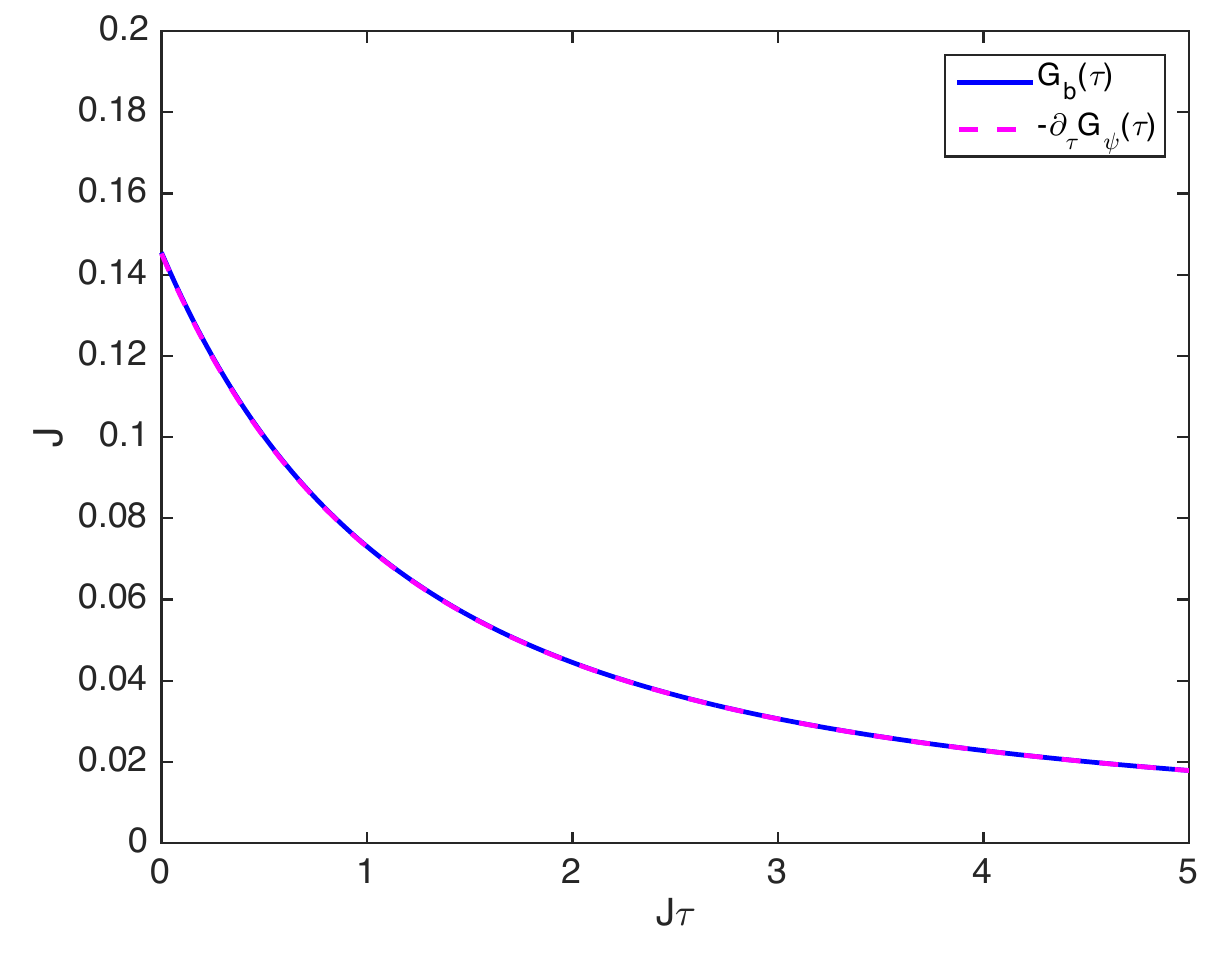}
\caption{Imaginary time Green's function at $T=0$ for $N=24$ Majorana fermions averaged over 100 samples. The blue solid line is $G_{bb}(\tau)$, and the pink dashed line is $-\partial_{\tau}G_{\psi}(\tau)$. }
\label{fig:dGpsi}
\end{figure}

An important purpose of the numerical study was to examine whether supersymmetry was unbroken in the $N \rightarrow \infty$
limit. In Fig.~\ref{fig:dGpsi} we test the basic relationship in Eq.~(\ref{SUSYequation}) between the fermion and boson Green's functions.
The agreement between the boson Green's function and the time derivative of the fermion Green's function is evidently excellent.
 
\begin{figure}[h]
\center
\includegraphics[width=3.5in]{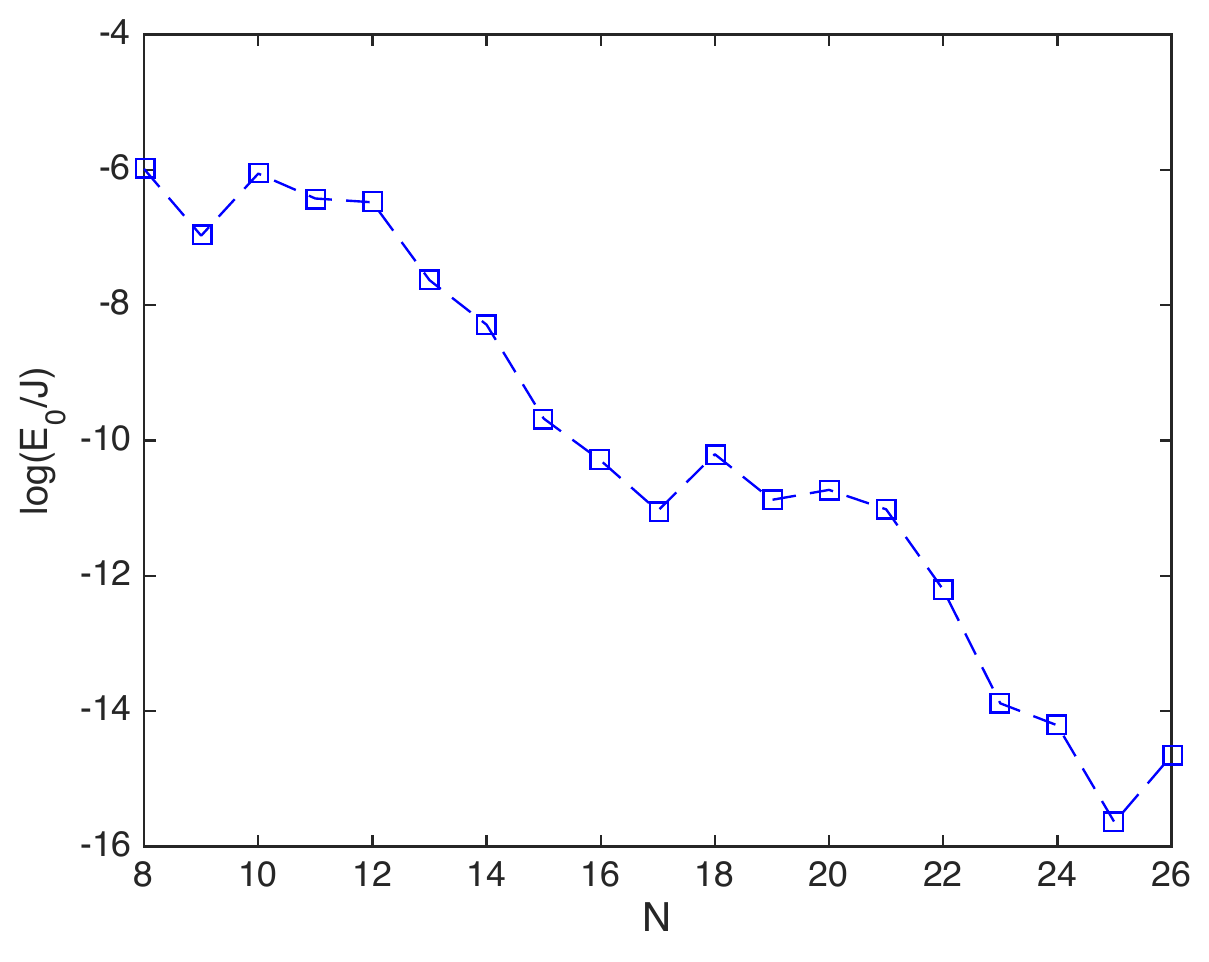}
\caption{Ground state energy as a function of $N$ in a log-linear plot, where we have averaged over 100 samples. 
The plot is compatible with an exponential decrease of $E_0$ with $N$. Notice also the structure in $E_0$ dependent on $N$ (mod 8).}
\label{E0}
\end{figure}

We also computed  the value of the ground state energy $E_0=\langle 0|QQ|0\rangle$. Supersymmetry is unbroken if an only if $E_0 =0$. 
We have found that $E_0$ is non-zero in the exact theory, but it becomes very small  for large  $N$. Indeed 
 fig.~\ref{E0} shows that $E_0$ does become very small,
and the approach to zero is compatible with an exponential decrease of $E_0$ with $N$.
 This is then compatible with a supersymmetric large $N$ solution, supersymmetry is then broken non-perturbatively in the $1/N$ expansion. 
 The combination of Figs.~\ref{fig:dGpsi} and~\ref{E0} is strong numerical evidence for the preservation of supersymmetry 
in the $N \rightarrow \infty$ limit (with suppersymmetry breaking at finite $N$). The ground state energy can be fitted well by $ E_0 \propto e^{ - \alpha S_0 } $
with $\alpha = 1.9 \pm .2 $, which is compatible with $\alpha =2$. Here $S_0 $ is the ground state entropy, \nref{GSEntropy} . 
This is smaller that the naive estimate for the interparticle level spacing which is $e^{ - S } $. 

Note that the breaking of supersymmetry is also compatible with the Witten index of this model which is $Tr[(-1)^F]=0$. This can be 
easily computed in the free theory. For $N$ odd we defined the Hilbert space by adding 
 an extra Majorana mode that is decoupled from the ones appearing in the Hamiltonian.

As in   Ref.~\cite{YLX16}, we found a ground state degeneracy pattern that depended upon $N$ (mod 8). The pattern in our case is (for $N\geq 3$)
\beq
\begin{tabular}{c|cccccccc}
\hline
  $N$ (mod 8) & 0 & 1 & 2 & 3 & 4 & 5 & 6 & 7 \\
 \hline
  Degeneracy  &2 & 4 & 4 & 4 & 4 & 4 & 2 & 2\\
\hline
\end{tabular}.
\eeq
For odd $N$ this degeneracy includes all the states in the Hilbert space defined by adding an extra decoupled fermion. 
We also found that the value of $E_0$ has structure dependent upon $N$ (mod 8), as is clear from Fig.~\ref{E0}.

\subsection{Scaling}

\begin{figure}[h]
\center
\includegraphics[width=3.3in]{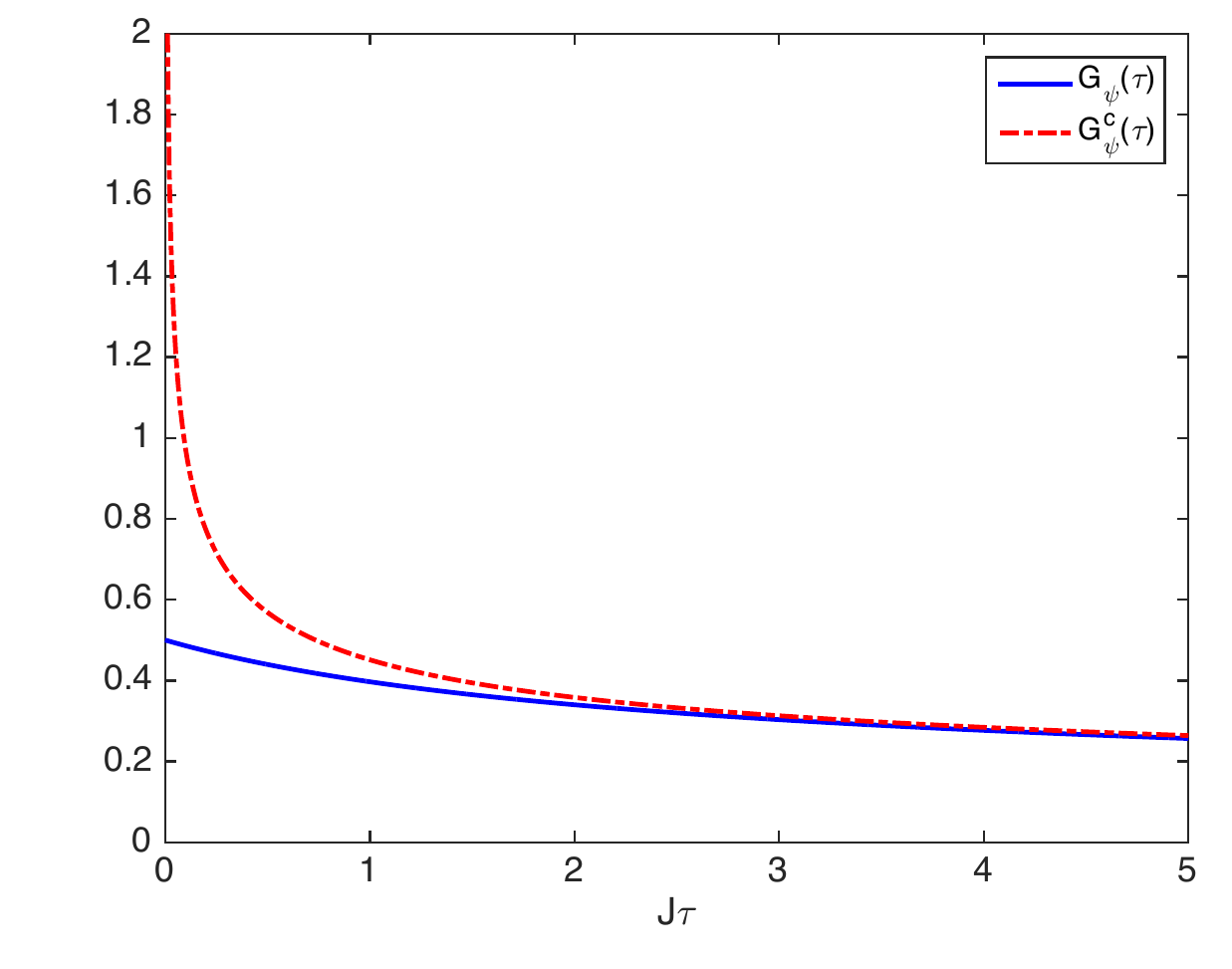}
\includegraphics[width=3.3in]{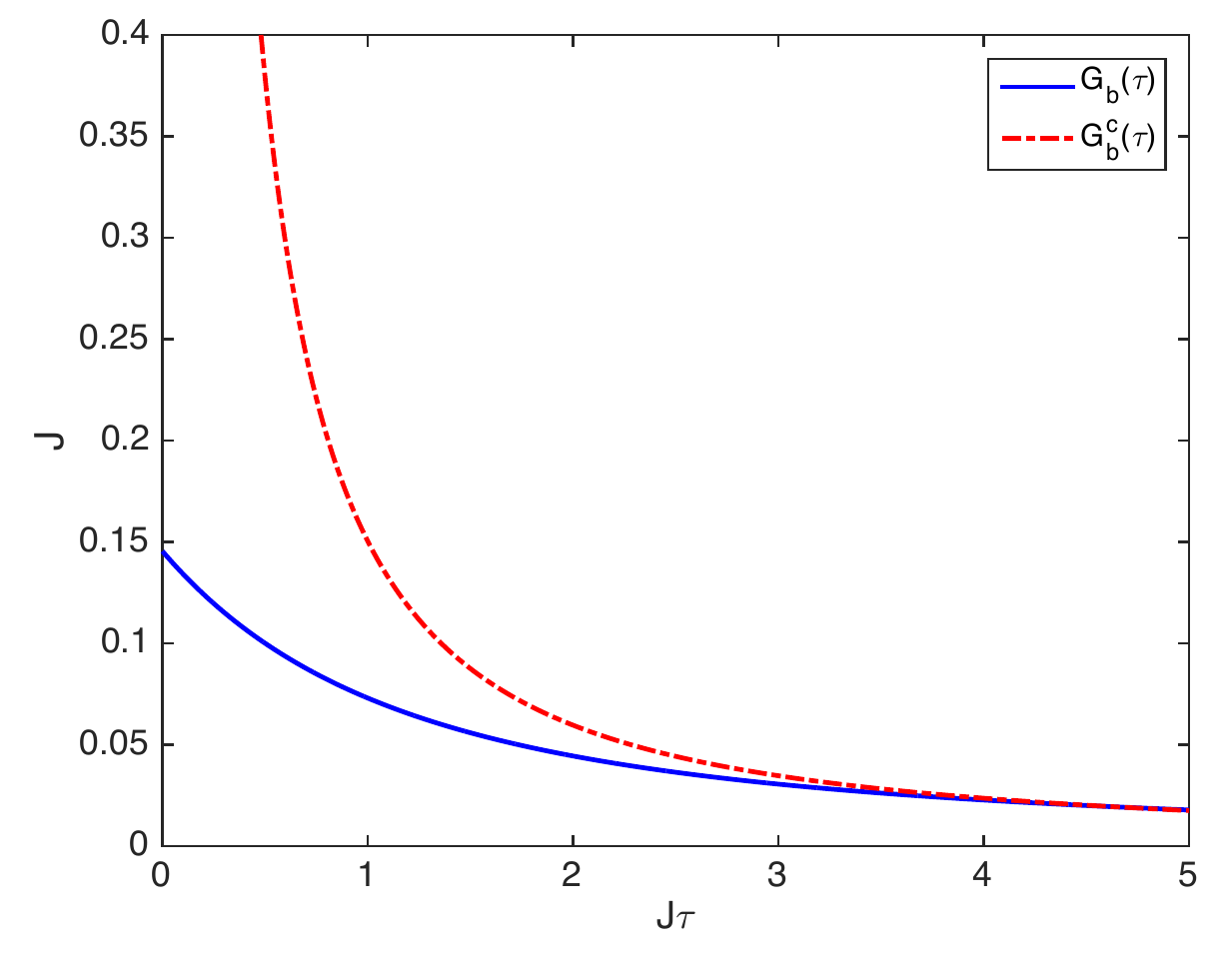}
\caption{Imaginary time Green's function at $T=0$ for $N=24$ Majorana fermions averaged over 100 samples. Left panel: blue solid line is $G_{\psi \psi}(\tau)$, red dotted-dashed line is the conformal solution $G_{\psi \psi}^c(\tau)$ in Eq.~(\ref{Gbc}); right panel: blue solid line is $G_{bb}(\tau)$, red dotted-dashed line is the conformal solution $G_{bb}^c(\tau)$ in Eq.~(\ref{Gbc}). }
\label{fig:GGc}
\end{figure}

We also compared our numerical results for the Green's functions with the conformal scaling structure expected at
long times and low temperatures. From Eq \nref{Jbb}, with $\hat q =3 $,  we expect that at $T=0$  and  large $\tau$
\beq
G_{\psi \psi}^c(\tau)=\frac{\mbox{sgn}(\tau)}{(2 \pi \sqrt{3})^{1/3}} |   J \tau|^{-1/3} \quad , \quad
G_{bb}^c(\tau)=\frac{1}{3 (2 \pi\sqrt{3})^{1/3}} J |J \tau|^{-4/3} . \label{Gbc}
\eeq
Fig~\ref{fig:GGc} shows that Eq.~(\ref{Gbc}) is obeyed well for large $J \tau$.

\begin{figure}[h]
\center
\includegraphics[width=3.4in]{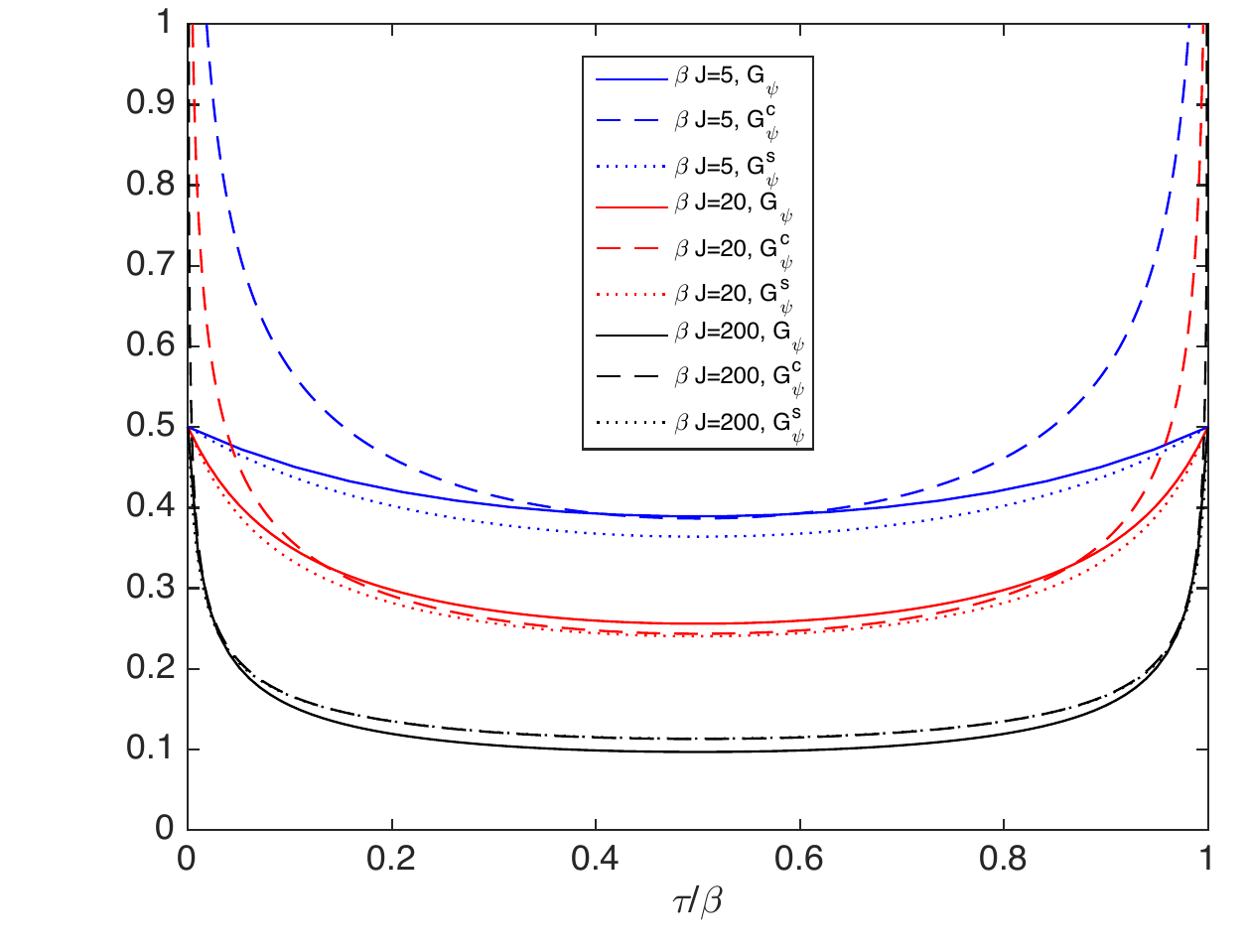}
\includegraphics[width=3.4in]{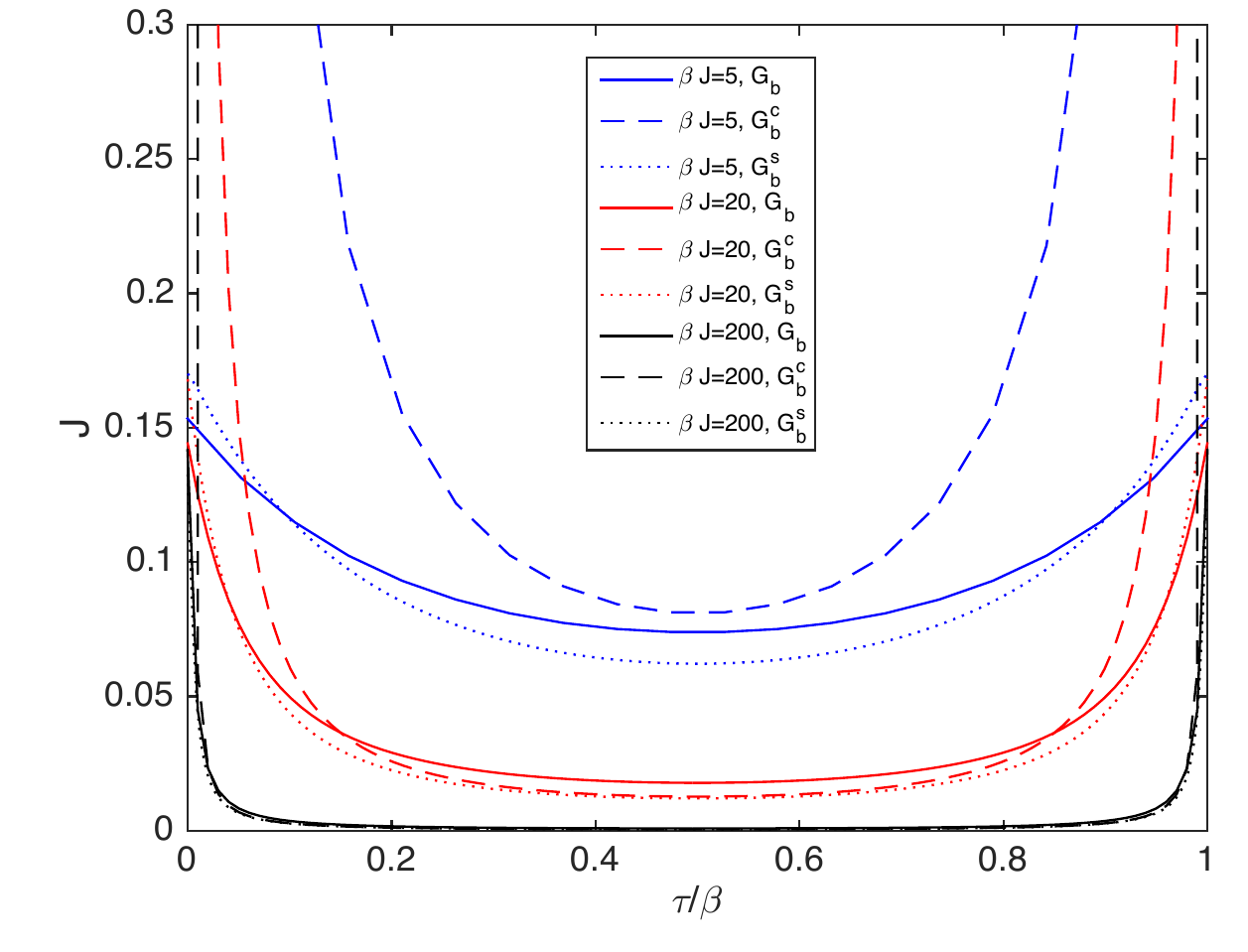}
\caption{Imaginary time Green's function at finite temperature for $N=20$ Majorana fermions averaged over 100 samples. Left panel is $G_{\psi\psi}(\tau)$ while right panel is $G_{bb}(\tau)$. Solid lines are the exact diagonalization result; dashed lines are conformal results as in Eq.~(\ref{GpsiGb}); dotted line are large $N$ result by numerically solving Eq.~(\ref{EOM12}) and Eq.~(\ref{EOM34}). Different colors correspond to different interaction strength: blue one is $\beta J=5$; red one is $\beta J=20$ and black one is $\beta J=200$. }
\label{Fig:GpsiGbfiniteT}
\end{figure}

We also extended this comparison to $T>0$, where we expect the generalization of Eq.~(\ref{Gbc}) to 
\beq
G_{\psi}^c(\tau) = \frac{\mbox{sgn}(\tau)}{(2 \pi \sqrt{3})^{1/3}} \left[\frac{\pi T}{J \sin{(\pi\tau T)}}\right]^{1/3}
\quad,\quad
G_{bb}^c(\tau) = \frac{1}{3(2 \pi \sqrt{3} )^{1/3}} J \left[\frac{\pi T}{ J \sin{(\pi\tau T)}}\right]^{4/3}
\label{GpsiGb}
\eeq
The comparison of these results with the numerical data appears in Fig.~\ref{Fig:GpsiGbfiniteT}.

\section{Superspace and super-reparameterization} \label{sec:superspace}

So far, we have seen that the main consequence of supersymmetry was the relationship Eq.~(\ref{SUSYequation})
between the boson and fermion Green's functions at $T=0$. However, as is clear from Eq.~(\ref{GpsiGb}), this simple relationship does
not extend to $T>0$. Of course, this is not surprising, since finite temperature breaks supersymmetry. 

Previous work on the SYK models has highlighted reparameterization and conformal symmetries \cite{PG98,kitaev2015talk,SS15,JMDS16} 
which allow one to map zero and non-zero temperature correlators.
This section will describe how supersymmetry and reparameterizations combine to yield super-reparameterization symmetries,
and the consequences for the correlators. 

As in the SYK model, most of this super-reparameterization symmetry is spontaneously broken. 
There is, however, a part of it that is left unbroken by \nref{GpsiGb}. This unbroken part includes both a bosonic $SL(2,R)$ group as well 
as two fermionic generators, giving an $OSp(1|2)$ global super-conformal group. 
These super-symmetry generators are emergent, and are different from the original supersymmetry
of the model. In particular, they square to general conformal transformations of the thermal circle rather than time translations. 
We will come back to this point more explicitly later.  

\subsection{Superspace}

Superspace offers a simple way to package together the degrees of freedom and equations of motion for Eq.~(\ref{SuperSYKH})
while making supersymmetry manifest. Concretely, we define a super-field
\begin{equation}
\Psi(\tau, \theta) = \psi(\tau) + \theta b(\tau)
\end{equation}
which is a function of both time and an auxiliary anticommuting variable $\theta$. 

Supersymmetry transformations combine with translations into a group of super-translations
\begin{equation} \la{SusyTr}
\tau \to \tau' = \tau + \epsilon + \theta \eta \qquad \qquad \theta \to \theta' = \theta + \eta
\end{equation}
It is well-known that a Grassman integral of the form 
\begin{equation}
\int d \theta d \tau F(\theta, \tau) 
\end{equation}
for some function of $\theta$ and $\tau$ is invariant under super-translation: if we expand $F(\theta', \tau') = F_1(\tau') + \theta' F_2(\tau')$ then, by 
definition, 
\begin{equation}
\int d \theta' d \tau' F(\theta', \tau') = \int d\tau' F_2(\tau') 
\end{equation}
and we find that 
\begin{equation}
\int d \theta d \tau F(\theta, \tau)  = \int d \theta d \tau F(\theta + \eta, \tau + \epsilon + \theta \eta) = \int d \tau \left(F_2(\tau + \epsilon)+ \eta \partial_\tau F_1(\tau + \epsilon)  \right)
\end{equation}
are the same up to total derivatives. 

The Lagrangian Eq.~(\ref{Leff}) can be written in a manifestly supersymmetric form
\beq
\mathcal{L}=\int d\theta(-\frac{1}{2}\Psi^iD_{\theta}\Psi^i- i C_{ijk}\Psi^i\Psi^j\Psi^k) \label{superL}
\eeq
by introducing the super-derivative operator 
\begin{equation}
D_\theta \equiv \partial_\theta + \theta \partial_\tau \qquad \qquad D_\theta^2 = \partial_\tau
\end{equation}
which is invariant under super-translations. Indeed, 
\begin{equation}
D_\theta F(\tau + \epsilon + \theta \eta,\theta + \eta) = \partial_{\theta'} F + (\theta + \eta) \partial_{\tau'} F = D_{\theta'} F(\tau', \theta')
\end{equation}

We can now derive the super equations of motion. Let us define
\beq
\mathcal{G}(\tau_1,\theta_1;\tau_2,\theta_2)=\langle \Psi(\tau_1,\theta_1)\Psi(\tau_2,\theta_2)\rangle
\eeq
This super-field includes both the bosonic bilinears $G_{\psi \psi}$ and $G_{bb}$ and the fermionic bilinears $G_{b \psi}$ and $G_{\psi b}$. 
The equations of motions of the disorder-averaged Lagrangian $\mathcal{L}$ can now be expressed in a manifestly supersymmetric way as
\beq 
D_{\theta_1} \mathcal{G}(\tau_1,\theta_1;\tau_3,\theta_3)+\int d\tau_2d\theta_2\mathcal{G}(\tau_1,\theta_1;\tau_2,\theta_2)\big(J \mathcal{G}(\tau_2,\theta_2;\tau_3,\theta_3)^2\big)=(\theta_1-\theta_3)\delta(\tau_1-\tau_3)\label{EOM}
\eeq

The right hand side is the supersymmetric generalization of the delta function:
\begin{equation}
F(\theta_1, \tau_1) (\theta_1-\theta_2)\delta(\tau_1-\tau_2) = F(\theta_2, \tau_2) (\theta_1-\theta_2)\delta(\tau_1-\tau_2)
\end{equation}

Some useful super-translation invariant combinations are $\theta_1 - \theta_2$ and $\tau_1 - \tau_2 - \theta_1 \theta_2$, 
which satisfies $D_1(\tau_1 - \tau_2 - \theta_1 \theta_2) = \theta_1 - \theta_2$. 
In a translation-invariant, supersymmetric vacuum of definite fermion number, the solution must take the form
\beq
\mathcal{G}(\tau_1,\theta_1;\tau_2,\theta_2)=G_{\psi \psi}(\tau_1-\tau_2)+\theta_1\theta_2G_{bb}(\tau_1-\tau_2)=G_{\psi \psi}(\tau_1-\tau_2-\theta_1\theta_2)
\eeq
If we use a vacuum that does not have definite fermion number, supersymmetry imposes that $G_{\psi b}=G_{b \psi}$, so in a translation-invariant, supersymmetric vacuum (without definite fermion number) we have
\bea
\mathcal{G}(\tau_1,\theta_1;\tau_2,\theta_2)&=&G_{\psi \psi}(\tau_1-\tau_2)+\theta G_{b \psi}(\tau_1-\tau_2)-\theta_2 G_{\psi b}(\tau_1-\tau_2)+\theta_1\theta_2G_{bb}(\tau_1-\tau_2)\nonumber\\
&=&
G_{\psi \psi}(\tau_1-\tau_2-\theta_1\theta_2)+(\theta_1-\theta_2)G_{b \psi}(\tau_1-\tau_2-\theta_1\theta_2)
\eea

Of course, the whole derivation of the effective action can be re-cast in superspace, starting from 
\beq
S_{\mathrm{eff}}=\int d\theta d \tau (-\frac{1}{2}\Psi^iD_{\theta}\Psi^i) + \frac{J}{3 N^2}\int d\theta_1 d \tau_1d\theta_2 d \tau_2 \left(\Psi^i \Psi^i \right)^3
\eeq
and introducing Lagrange multipliers $\Sigma(\theta_1, \tau_1;\theta_2, \tau_2) = \Sigma_{bb}(\tau_1, \tau_2) + \cdots$. 

One point to note is that this effective action contains also the fermionic 
bilinears $G_{\psi b}$, $G_{b\psi}$, which are important for making the action supersymmetric. Of course, such terms are also important when we compute 
correlation functions, as will be done in section \ref{4pt}. These terms can be consistenly set to zero when we consider the classical equations, as was done in 
section \ref{sec:eff}. 

\subsection{Super-reparameterization}

We now turn to a discussion of the reparameterization symmetry, discussed previously \cite{PG98,kitaev2015talk,SS15,JMDS16}
for the non-supersymmetric SYK model.

If we drop the first term, the supersymmetric equations (\ref{EOM}) have a large amount of symmetry: general coordinate transformations 
\begin{equation}
\tau \to \tau'(\tau, \theta) \qquad \qquad \theta \to \theta'(\tau, \theta)
\end{equation}
accompanied by a re-scaling 
\begin{equation}
\mathcal{G}(\tau_1,\theta_1;\tau_2,\theta_2) = \mathrm{Ber} (\theta'_1, \tau'_1,\theta_1, \tau_1)^{\frac13} \mathrm{Ber} (\theta'_2, \tau'_2,\theta_2, \tau_2)^{\frac13} \mathcal{G}(\tau'_1,\theta'_1;\tau'_2,\theta'_2)
\end{equation}
where the {\it Berezinian} 
\begin{equation}
\mathrm{Ber} (\theta', \tau',\theta, \tau) \equiv \mathrm{Ber} \begin{pmatrix}\partial_\tau \tau' &  \partial_\tau \theta'  \cr \partial_\theta \tau' & \partial_\theta \theta'  \end{pmatrix} 
\end{equation}
is a generalization of the Jacobian which encodes the change in the measure $d\theta d\tau$ and in the supersymmetric delta function. 

These transformations generalize the usual re-parameterization symmetry of the standard SYK model. They include two bosonic and two fermionic 
functions of $\tau$. The second bosonic generator is a generalization of the scaling symmetry (\ref{symNai}) and we expect it to be broken by the UV boundary conditions. 
More precisely, we can Taylor expand 
\begin{equation}\label{eq:taylor}
\tau'_1 - \tau'_2 - \theta'_1 \theta'_2 = (\theta_1 - \theta_2) \left( D_{\theta_2} \tau_2' - \theta_2' D_{\theta_2} \theta_2'\right) + (\tau_1 - \tau_2 - \theta_1 \theta_2) \left[  \partial_{\tau_2} \tau_2' - \theta_2' \partial_{\tau_2} \theta_2'+ \cdots \right]
\end{equation}
where the ellipsis indicate higher order terms. 

We observe that the short-distance singular behaviour of $\mathcal{G}(\tau_1,\theta_1;\tau_2,\theta_2)$ will only be preserved if 
the coordinate transformations satisfy 
\begin{equation} \label{eq:cond}
D_{\theta} \tau' = \theta' D_{\theta} \theta'
\end{equation}
and furthermore the square of the Berezinian factors coincide with the coefficient of $(\tau_1 - \tau_2 - \theta_1 \theta_2)$ in (\ref{eq:taylor}),
which simplifies to $(D_\theta \theta')^2$ thanks to (\ref{eq:cond}).

These constraints define a well known set of transformations: super-reparameterizations. \footnote{The invariance of the equations of motion 
under the group of general coordinate transformations, rather than super-reparameterizations only, was noticed independently by E. Witten 
after we submitted an earlier version of this paper.} We will now review their basic properties and discuss their implications 
for the low-energy physics. 

The supersymmetric generalization $\mathrm{SDiff}$ of the reparameterization group $\mathrm{Diff}$
can be defined as the set of coordinate transformations $(\tau, \theta) \to (\tau', \theta')$ on the super-line which preserve the super-derivative 
$D_\theta$ up to a super-Jacobian factor $D_\theta \theta'$:
\begin{equation}
D_\theta = D_\theta \theta' D_{\theta'}
\end{equation}

The bosonic part of $\mathrm{SDiff}$ is the usual diffeomorphism group $\mathrm{Diff}$, 
acting as 
\begin{equation}
\tau \to \tau' = f(\tau)  \qquad \qquad \theta \to \theta' = \sqrt{\partial_\tau f} \theta,
\end{equation}
where $f(\tau)$ is the usual reparameterization. Indeed, $D_\theta \theta' = \sqrt{\partial_\tau f}$ and 
\begin{equation}
D_\theta F( f(\tau),\sqrt{\partial_\tau f} \theta ) = \sqrt{\partial_\tau f} \partial_{\theta'} F + \theta \partial_\tau f \partial_{\tau'} F = \sqrt{\partial_\tau f} D_{\theta'} F
\end{equation}

In general, 
\begin{equation}
D_\theta F(\tau',\theta') = D_\theta \tau' \partial_\tau' F + D_\theta \theta' \partial_\theta' F = D_\theta\theta' D_{\theta'} F + \left(D_\theta \tau' - \theta' D_\theta\theta'  \right)\partial_{\tau'} F
\end{equation}
and thus super-reparameterizations are coordinate transformations constrained by 
\begin{equation}
D_\theta \tau' = \theta' D_\theta\theta' 
\end{equation}

Infinitesimally, the  super-reparameterizations, generated by a bosonic function $\epsilon(\tau)$ and a fermionic function $\eta(\tau)$, are 
\begin{equation}
\delta \tau = \epsilon(\tau) + \theta \eta(\tau) \qquad \qquad \delta \theta = \eta(\tau) + \frac{\theta}{2} \partial_\tau \epsilon(\tau) 
\end{equation}
A useful parameterization of finite transformations is
\begin{equation} \la{susyvar}
\tau' = f(\tau + \theta \eta(\tau) ) \qquad \qquad \theta' = \sqrt{\partial_\tau f(\tau)}\left[ \theta  + \eta(\tau) + \frac12 \theta \eta(\tau)\partial_\tau \eta(\tau)\right].
\end{equation}
This is just the composition of a general fermionic transformation of parameter $\eta$ followed by a diffeomorphism.  The original supersymmetry transformation 
\nref{SusyTr} acts in these variables as 
\be \la{SusyComp} 
f \to f + f' \epsilon \eta ~,~~~~~~~~ \eta \to \eta + \epsilon + \epsilon \eta' \eta 
\ee

Finally, we note that
global super-conformal transformations are generated by super-translations and the inversion 
\begin{equation}
\tau \to \tau' =-\frac{1}{\tau}  \qquad \qquad \theta \to \theta' = \frac{\theta}{\tau} 
\end{equation}
They form an OSp$(1|2)$ group with three bosonic generators and two fermionic generators. 
These are fractional linear transformations 
\begin{equation}
\tau' = \frac{a \tau + \alpha \theta + b}{c \tau + \gamma \theta + d} \qquad \qquad \theta' = \frac{\beta \tau + e \theta + \delta}{c \tau + \gamma \theta + d} 
\end{equation}
with coefficients subject to appropriate quadratic constraints:
\begin{equation}
(\beta \tau + e \theta + \delta)(e + \theta \beta)+ (a \tau + \alpha \theta + b)(- \gamma + \theta c) - (c \tau + \gamma \theta + d)(- \alpha + \theta a) =0 
\end{equation}
i.e. 
\begin{equation}
e\beta - a  \gamma+ \alpha c =0 \qquad  e^2 +\beta \delta+2 \alpha \gamma+ b c-a d =0 \qquad e \delta -\gamma b + \alpha d =0 
\end{equation}
Choosing an arbitrary overall scale for the coefficients we can also write this as 
\begin{equation}
\begin{pmatrix}e & -\alpha & -\gamma \cr \beta & a & c \cr \delta & b & d \end{pmatrix}\begin{pmatrix}1 & 0 & 0 \cr 0&0&-1 \cr 0&1&0 \end{pmatrix}   \begin{pmatrix}e & \beta& \delta \cr \alpha & a & b \cr \gamma & c & d \end{pmatrix}= \begin{pmatrix}1 & 0 & 0 \cr 0&0&-1 \cr 0&1&0 \end{pmatrix}  
\end{equation}

We will now show that the Berezinian factor involved the super-reparameterization symmetry of the equation of motion in Eq.~(\ref{EOM}) 
without the first derivative term can be simplified to the super-Jacobian factor and thus the symmetries are compatible with the UV boundary conditions. 

The result follows from a basic fact about superspace integrals: under super-reparameterizations, 
\begin{equation}
\int d \tau' d \theta' F(\tau', \theta')  = \int d \tau d \theta D_\theta \theta' F\left(\tau'(\tau, \theta), \theta' (\tau, \theta)\right) 
\end{equation}
Indeed, the integration measure changes by the Berezinian 
\begin{align}
\mathrm{Ber} \begin{pmatrix}\partial_\tau \tau' &  \partial_\tau \theta'  \cr \partial_\theta \tau' & \partial_\theta \theta'  \end{pmatrix} 
&= \mathrm{Ber} \begin{pmatrix}\partial_\tau \tau'  & \partial_\tau \theta'  \cr D_\theta \tau' - \theta \partial_\tau \tau' &  D_\theta \theta' - \theta \partial_\tau \theta' \end{pmatrix} =\mathrm{Ber} \begin{pmatrix}\partial_\tau \tau'  & \partial_\tau \theta'  \cr D_\theta \tau'  &  D_\theta \theta' \end{pmatrix} \cr
&= (D_\theta\theta')^{-1} \mathrm{Ber} \begin{pmatrix}\partial_\tau \tau'  & \partial_\tau \theta'  \cr \theta'  &  1\end{pmatrix}= (D\theta')^{-1}\left( \partial_\tau \tau'  - \partial_\tau \theta' \theta'  \right) \cr
&= (D_\theta\theta')^{-1}\left( D_\theta^2 \tau'  - D_\theta^2 \theta' \theta'  \right) = D_\theta\theta'
\end{align}

That means that we can make the equations of motion and effective actions invariant under $\mathrm{SDiff}$ 
as long as we transform
\begin{equation}
\mathcal{G}(\tau_1,\theta_1;\tau_2,\theta_2) = (D_{\theta_1} \theta'_1)^{\frac13} (D_{\theta_2} \theta'_2)^{\frac13} \mathcal{G}(\tau'_1,\theta'_1;\tau'_2,\theta'_2)
\end{equation}
The power becomes $\frac{1}{\hat q}$ for the generalized model. 

This is our proposal for the IR symmetries of the equations of motion 
\beq 
D_\theta \mathcal{G}(\tau,\theta ;\tau'',\theta'' )+\int d\tau'd\theta' \mathcal{  G}(\tau,\theta ;\tau',\theta' )\big(J \mathcal{G}(\tau',\theta';\tau'',\theta''   )^2\big)=(  \theta-  \theta'')\delta(\tau   -\tau''  )\label{EOMN1}
\eeq

Under bosonic reparameterizations, $G_{\psi \psi}$ and $G_{b b}$ transform independently, with the expected weight. 
The fermionic generators, though, mix $G_{\psi \psi}$ and $G_{bb}$ with $G_{\psi b}$ and $G_{b \psi}$.


\subsection{Super-Schwarzian}
\label{SuperSchwarzian}

For the non-supersymmetric SYK model, following a proposal by Kitaev \cite{kitaev2015talk}, Maldacena and Stanford \cite{JMDS16}
showed that the fluctuations about the large $N$ saddle point are dominated by a near-zero mode associated with reparameterizations
of the Green's function, and the action of the this mode is the Schwarzian. Here, we generalize this structure 
to the supersymmetric case. SuperSchwarzians have been previously discussed in \cite{Friedan:1986rx,Cohn:1986wn}.

The Schwarzian derivative $S[f(\tau),\tau]$ is a functional of the reparameterization $f(\tau)$ which vanishes if 
$f(\tau)$ is a global conformal transformation. A direct way to produce $S[f(\tau),\tau]$ is to consider 
the expression 
\begin{equation}
\partial_{\tau_1} \partial_{\tau_2} \log\frac{\tau'_1 - \tau'_2}{\tau_1 - \tau_2} = \frac{\partial_{\tau_1} \tau_1'  \partial_{\tau_2} \tau_2'}{(\tau'_1 - \tau'_2)^2} - \frac{1}{(\tau_1 - \tau_2)^2}
\end{equation}
which vanishes if $\tau \to \tau'$ is a global conformal transformation. 
In the limit $\tau_2 \to \tau_1$ we recover (up to a factor of $6$) the usual Schwarzian
\begin{equation}
S[f(\tau),\tau] = { f'''\over f' } - { 3 \over 2 } \left( { f'' \over f' } \right)^2 
\end{equation}
This definition makes the chain rule manifest: 
\begin{equation}
S[g(f(\tau)),\tau] = \left(\partial_\tau f(\tau) \right)^2 S[g(f(\tau)),f(\tau)] + S[f(\tau),\tau] 
\end{equation}

The expression 
\begin{equation}
D_1 D_2 \log\frac{\tau'_1 - \tau'_2-\theta'_1 \theta'_2}{\tau_1 - \tau_2- \theta_1 \theta_2} = \frac{D_1 \tau_1'  D_2 \tau_2'}{\tau'_1 - \tau'_2-\theta'_1 \theta'_2} - \frac{1}{\tau_1 - \tau_2- \theta_1 \theta_2}
\end{equation}
vanishes when $(\tau', \theta')$ are obtained from $(\tau, \theta)$ by a global superconformal transformation.
This is evident for super-translations and easy to check for inversions. 
Taking another super-derivative $D_1$ and the limit $(\tau_1, \theta_1) \to (\tau_2, \theta_2)$ gives us 
the super-Schwarzian derivative 
\begin{equation}
S[\tau', \theta';\tau, \theta] = \frac{D^4 \theta'}{D \theta'} - 2 \frac{D^3 \theta' D^2 \theta'}{(D \theta')^2} = S_f(\tau', \theta';\tau, \theta) + \theta S_b(\tau', \theta';\tau, \theta)
\end{equation}
which satisfies a chain rule of the form 
\begin{equation}
S[\tau'', \theta'';\tau, \theta] = \left(D \theta' \right)^3 S[\tau'', \theta'';\tau', \theta'] + S[\tau', \theta';\tau, \theta] 
\end{equation}
The bosonic piece $S_b$ reduces to the usual Schwarzian derivative for standard reparameterizations. 
That means that the super-space action
\bea
& &- \int d \tau d \theta S[\tau', \theta';\tau, \theta] = -\int d\tau S_b(\tau', \theta';\tau, \theta)= 
\cr
&=&- \half \int dt S(f,\tau) + \eta \eta''' + 3 \eta' \eta'' - S(f,\tau) \eta \eta' \label{CompVar}
\eea
is a natural supersymmetrization of the Schwarzian action. In the second line we used  \nref{susyvar} to 
write the action in component fields. 
Infinitesimally, $f(\tau ) = \tau + \epsilon(\tau)$, we get 
$\frac{1}{4} (\epsilon'')^2 +   \eta'  \eta''$. 
And around the thermal solution, with $\beta = 2 \pi $,  $f(\tau) = \tan{ \tau + \epsilon(\tau) \over 2 } $, we get 
$ \frac{1}{4} ( {\epsilon'' }^2 - \epsilon'^2 ) +   \eta'  \eta'' - { 1\over 4} \eta \eta' $.
This contains solutions with the expected time dependence to be associated to the generators 
of the superconformal group. 
The bosonic ones as as in \cite{JMDS16}. The fermion zero modes have a behavior $\eta \sim e^{ \pm i \tau/2}$ (or 
$\eta \sim e^{ \pm i \pi \tau/\beta }) $. 

The action of supersymmetry on these variables \nref{SusyComp} would suggest that supersymmetry is always broken since $\eta$ shifts under 
supersymmetry as a goldstino. More explicitly a configuration that preserves supersymmetry is a solution that is left invariant under \nref{SusyComp}. 
For example, consider   the configuration $f =\tau$ and $\eta=0$, which is the zero temperature solution and is expected to be invariant under supersymmetry. 
But we see that this is not the case since \nref{SusyComp} shows that the transformation leads to a non-zero value of $\eta$. 
However, it is possible   to combine this supersymmetry
with one of the $OSp(1|2)$ transformations, which acts as a super translation on $t',\theta'$ so as to cancel this term and leave the solution invariant. Thus, the
$f=\tau$, $\eta=0$  solution is invariant under supersymmetry. 
On the other hand, when we expand around the thermal solution, it is no longer possible to cancel the
supersymmetry variation of $\eta$ at all points on the thermal circle. So supersymmetry is broken in this case. 
A similar issue arises with ordinary translations, under $\tau \to \tau + b$. The solution $f=\tau$ is not invariant. On the other hand, if we combine
this translation with one of the $SL(2,R)$ transformations $f \to f -b$, then we find that the combination of the two leave the solution invariant.

Notice that even though the original supersymmetry of the model is broken by the finite temperature, the low energy configuration is
invariant under a global $OSp(1|2)$  subgroup of all super-reparameterizations. These transformations involve also fermionic generators, under full rotations along the 
thermal circle, these generators pick up a minus sign, compatible with the fermionic boundary conditions on the circle\footnote{
 This is conceptually similar to the way in 
which a 1+1 dimensional supersymmetric CFT preserves supersymmetry in the NS sector. The preserved supercharges have non-zero energy and momentum. }. 
The situation is somewhat similar to the purely bosonic case, where the finite temperature breaks the scaling symmetry in physical time, but we 
still have a symmetry of correlators under a full $SL(2)$ symmetry, the symmetry leaving the Schwarzian invariant.

These zero modes are unphysical and should not be viewed as degrees of freedom of the model. In particular, when we 
compute the one loop determinant for fluctuations around the classical large $N$ solution, their absence from the path integral, gives 
an interesting $\beta J$ dependence to the low temperature partition function ($ 1\ll \beta J \ll N $) 
\be  \label{ZeroMOne}
 Z_{\rm 1-loop} \sim { \beta J \over (\beta J )^{3/2} } e^{ S_0 + c/(2 \beta J) }  \longrightarrow \rho(E) \sim { 1 \over \sqrt{ E J } } e^{ S_0 + \sqrt{ 2 c E/J}}
\ee
The denominator comes from the three bosonic zero modes and the numerator from the fermionic 
ones\footnote{The net prefactor of $\beta^{-1/2}$ in \nref{ZeroMOne} implies that the partition function 
$|Z(\beta + i  t)|$ should go like $ t^{-1/2}$ for large times in the ``slope'' regime in \cite{StanfordGroup}. 
Numerically we found that the ``slope'' is $-0.54\pm 0.08$ in a regime which is naively outside the 
regime of validity of our derivation of \nref{ZeroMOne}, which can be viewed as an indication that
perhaps \nref{ZeroMOne} would not receive corrections, as in the purely bosonic case \cite{StanfordGroup}. }. Here $S_0$ is the ground state entropy and the
temperature independent contribution to the one loop partition function and the term $c/(2 \beta J)$ is the contribution to the free energy coming from the 
Schwarzian action ($c$ is of order $N$).
 We have also indicated the implication for the density of states, which is obtained by integrating over $\beta$ (along a suitable contour), considering
both the saddle point contribution as well as the gaussian integral around the saddle.

\section{$\mathcal{N}=2$ supersymmetry}
\label{sec:N2}

This section turns to the generalization to $\mathcal{N}=2$ supersymmetry. The real fermion $\psi^i$ is replaced by
complex fermions $\psi^i$ and $\bar \psi_i$, and the
supercharge $Q$ in Eq.~(\ref{defQ}) is replaced by a pair of charges $Q$ and $\bar Q$. The defining relations are
\bea
\{\psi^i , \bar \psi_j\} &=& \delta^i_j \quad, \quad \{\psi^i ,  \psi^j \} = 0 \quad, \quad \{\bar \psi_i ,  \bar \psi_j \} = 0 \nn
Q &=& i \sum_{1\leq i < j< k\leq N} C_{ijk} \psi^i \psi^j \psi^k \nn
\bar Q &=& i \sum_{1\leq i < j< k\leq N} \bar C^{ijk} \bar \psi_i \bar \psi_j \bar \psi_k
\eea
which imply $Q^2 = \bar Q^2 = 0$.
The theory has a U(1)$_{\rm R}$ R-symmetry, under which the fermions
$\psi^i$ and $\bar \psi_i$ carry charges $1/3$ and $-1/3$. As is customary, we normalize the $U(1)_R$ charge so that the supercharges carry charge $\pm 1$. 
The supersymmetry acts on the fermionic variables as 
\begin{equation}
[Q, \psi^i ] =0 \qquad \qquad [Q, \bar \psi_i] =  \bar b^i \equiv i \sum_{1\leq j < k\leq N} C_{ijk} \psi^j \psi^k
\end{equation}

The Hamiltonian replacing Eq.~(\ref{SuperSYKH}) is now 
\beq
\mathcal{H} = \{Q, \bar Q\} =|C|^2 + \sum_{i,j,k,l} J_{ij}^{kl} \psi^i \psi^j \bar \psi_k \bar \psi_l \label{N2H}
\eeq
We note this Hamiltonian has the same form as the complex SYK model introduced in Ref.~\cite{SS15}, 
but now the complex couplings $J_{ij}^{kl}$ are not independent random variables. Instead we take the $C_{ijk}$ to
be independent random complex numbers, with the non-zero second moment
\beq
\overline{ C_{ijk} \bar C^{ijk}} = \frac{2 J 
}{N^2}
\eeq
replacing Eq.~(\ref{CCaverage}).

The subsequent analysis of Eq.~(\ref{N2H}) closely parallels the $\mathcal{N}=1$ case.
The main difference is that we now introduce complex non-dynamical auxiliary bosonic fields $b^i$ and $\bar b_i$ to linearize the supersymmetry transformations. 
The model can also be generalized so that $Q$ is built from products of  $\hat q $ fermions so that  the Hamiltonian involves  up to $ 2 \hat q -2 $ fermions. 

The equations of motion are a complexified version of the $\mathcal{N}=1$ equations. We will describe them momentarily. The fermion also has scaling dimension 
$\Delta_\psi = 1/( 2 \hat q )$ and R-charge $1/\hat q$. Notice that the R-charge of $\psi$ is twice its scaling dimension, 
which is as expected for a super-conformal chiral primary field. As is conventional the $U(1)_R$ charge is normalized so that the supercharge has charge one. 
  The $U(1)_R$ charge does not commute with the supercharges. There is however a $Z_{\hat q}$ group of this $U(1)$ symmetry that acts on the fermions as 
  $\psi^j \to e^{ 2 \pi i r\over \hat q} \psi^i$ which does leave the supercharge invariant and is a global symmetry commuting with supersymmetry. 
  Note that the quantization condition on the $U(1)_R$ charge, $Q_R$,  is that $ \hat q Q_R $ should be an integer.

This fact enables us to compute an simple generalization of the Witten index defined as 
\be \la{WittIn}
W_r = Tr[ (-1)^F e^{ 2\pi i r Q_R } ] = Tr[ (-1)^F  g^r ] = \left[ 1 - e^{ 2 \pi i r \over q } \right]^N = e^{ i N \pi ( {r \over \hat q}  - \half ) } \left[ 2 \sin { \pi r \over \hat q } \right]^N
\ee
where $g$ is the generator of the $Z_{\hat q }$ symmetry, and $Q_R$ is the $U(1)_R$ charge. 
In the third equality we have used that the index is invariant under changes of the coupling and computed it in the free theory, with $J=0$.   
The Witten index is maximal for $ r = { (\hat q \pm 1) / 2} $ where its absolute value is greatest and equal to 
\be
\log |W_{r= {\hat q \pm 1 \over 2 } } | = N \log[ 2 \cos { \pi \over 2 \hat q } ].
\ee
The right hand side happens to be the same as the value of the ground state entropy computed using the large $N$ solution, which is the same as 
\nref{GSEntropy}, up to an extra overall factor of two because now the fermions are complex. 
In general, these Witten indices should be a lower bound on the number of ground states, and also a lower bound on the large $N$ ground state entropy
(recall that in the ${\cal N}=1$ case we had that the Witten index was zero). The fact that the bound is saturated tells us that most of the states contributing
to the large $N$ ground state entropy are actually true ground states of the model. Thus, in this case supersymmetry is not broken by $e^{- N}$ effects. 

We have also looked at exact diagonalization of the theory, and computed the number of states for different values of the $U(1)_R$ charge. \footnote{Recall that the 
ground states of a quantum mechanics with $\mathcal{N}=2$ supersymmetry are in one-to-one correspondence with the cohomology of the $Q$ supercharge. 
This is easier to compute than the eigenvalues and eigenstates of the Hamiltonian.}
Let us define the R-charge so that it goes between $-N/\hat q \leq Q_R \leq N/\hat q$, in increments of $1/\hat q$. 
We have looked at the case $\hat q =3$ and we found the following degeneracies, $D(N,Q_R)$,  as a function of  $N$ and the charge
\bea
D(N,0 ) &=& 2 \, 3^{ N/2 -1} ~,~~~~~D(N,\pm { 1 \over 3 } ) =  3^{ N/2 -1} ~~,~~~~~~~~~{\rm for}~~ N ~~~{\rm even} 
\cr D(N, \pm { 1\over 6 }) &= & 3^{ (N-1)/2} ~,~~~~~~~~ {\rm for}~~ N =3\,{\rm mod}\,4
\cr
 D(N, \pm { 1\over 6 }) &= & 3^{ (N-1)/2} ~,~~~~~D(N,\pm { 3 \over 6 } ) = 1~{\rm or} ~~3 ~,~~~~~~~~ {\rm for}~~ N =1\,{\rm mod}\,4
\eea
And we have $D(N,Q_R) =0$ outside the cases mentioned above. Therefore, we see that the degeneracies are concentrated on states with very small values of
the $R$ charge. Of course, these values are consistent with the Witten index in \nref{WittIn} for $\hat q=3$. 

\subsection{Superspace and super-reparameterization}

Generalizing  previous discussions, we now expect that the fluctuations about the large $N$ saddle
point are described by spontaneously broken $\mathcal{N}=2$ super-reparameterization invariance, which includes a U(1)$_{\rm R}$ current algebra. 
The U(1)$_{\rm R}$ is similar to  the  emergent local U(1) symmetry that is present also for  for the non-supersymmetric complex SYK
  ~\cite{SS15}. In the low energy effective theory, this local symmetry  is broken down
to a global U(1), and so there is an associated gapless phase mode \cite{FGS16}.

Consider a $\mathcal{N}=2$ super-line, parameterized by a bosonic variable $\tau$ and a fermionic variables $\theta$ and $\bar \theta$. 
The super-translation group consists of the transformations
\begin{equation}
\tau \to \tau' = \tau + \epsilon + \theta \bar \eta + \bar \theta \eta  \qquad \qquad \theta \to \theta' = \theta + \eta \qquad \qquad \bar \theta \to \bar \theta' = \bar \theta + \bar \eta
\end{equation}
They preserve the super-derivative operators 
\begin{equation}
D \equiv \partial_\theta + \bar \theta \partial_\tau \qquad \qquad \bar D \equiv \partial_{\bar \theta} + \theta \partial_\tau
\end{equation}
Notice the super-translation invariant combination $\Delta_{12} = \tau_1 - \tau_2 - \theta_1 \bar \theta_2- \theta_2 \bar \theta_1$
which satisfies $D_1\Delta_{12}  = \bar \theta_1 - \bar \theta_2$ and $\bar D_1\Delta_{12}  = \theta_1 - \theta_2$. 
There is also an obvious  U(1)  symmetry rotating $\theta$ and $\bar \theta$ in opposite directions.  

We can package the complex fermions and scalars into a {\it chiral} superfield, i.e. a superfield $\Psi^i$ 
constrained to satisfy 
\begin{equation}
\bar D \Psi^i = 0 
\end{equation}
which is solved by 
\begin{equation}
\Psi^i(\tau, \theta, \bar \theta) = \psi^i(\tau + \theta \bar \theta)+ \theta b^i
\end{equation}
Notice that both the conjugate $\bar \Psi_i$ and $D \Psi^i$ are {\it anti-chiral}, i.e. are annihilated by $D$. 

The bi-linear $\mathcal{G} = \Psi^i \bar \Psi^i$ is thus {\it chiral-anti-chiral}, annihilated by $\bar D_1$ and by $D_2$. 
The equations of motion:
\beq 
D_\theta \mathcal{G}(\tau,\theta, \bar \theta ;\tau'',\theta'', \bar \theta'')+\int d\tau'd\theta' \mathcal{\bar G}(\tau,\theta,\bar \theta;\tau',\theta',\bar \theta')\big(J \mathcal{G}(\tau',\theta',\bar \theta';\tau'',\theta''\bar \theta'')^2\big)=(\bar \theta-\bar \theta'')\delta(\tau- \theta \bar \theta -\tau'' + \theta'' \bar \theta'')\label{EOMN2}
\eeq
are anti-chiral both in the first and the last set of variables. The equations involve the integral of chiral functions of the middle set of variables over the chiral measure $d\tau'd\theta'$ and is thus invariant under supersymmetry. Furthermore, 
even the delta function is  anti-chiral.

The analysis of re-parameterization invariance proceeds as before. 
If we consider a general coordinate transformation, we have 
\begin{equation}
D_\theta F(\tau',\theta', \bar \theta') = D_\theta \tau' \partial_\tau' F + D_\theta \theta' \partial_\theta' F + D_\theta \bar \theta' \partial_{\bar \theta'} F 
= D_\theta\theta' D_{\theta'} F + D_\theta \bar \theta' D_{\bar \theta'} F + \left(D_\theta \tau' - \bar \theta' D_\theta\theta'- \theta' D_\theta\bar \theta'  \right)\partial_\tau' F
\end{equation}
The $\mathcal{N}=2$ super-reparameterizations are coordinate transformations constrained by 
\begin{align}
D_\theta \bar \theta' &= 0 \qquad \qquad D_\theta \tau' = \bar \theta' D_\theta\theta'   \cr
D_{\bar \theta} \theta' &= 0 \qquad \qquad D_{\bar \theta} \tau' = \theta' D_{\bar \theta} \bar \theta'  
\end{align}
with super-Jacobian factor $D_\theta \theta'$.  

These transformations map chiral super-fields to chiral super-fields. The converse is also true. In particular, 
in the $\mathcal{N}=2$ case we do not have the freedom to do general coordinate transformations of $\tau$, $\theta$, $\bar \theta$ which would violate the chirality constraints on the 
superfields. Extra symmetries which generalize (\ref{symNai}) still appear, thought, and we will discuss them momentarily. 

The bosonic transformations, including re-parameterization and a position-dependent 
U(1) transformation, are 
\begin{align}
\theta' &= e^{i a(\tau)} \sqrt{\partial_\tau f(\tau)} \theta \cr
\bar \theta' &= e^{-i a(\tau)} \sqrt{\partial_\tau f(\tau)} \bar \theta \cr
\tau' &= f(\tau) 
\end{align}
There are also chiral and anti-chiral fermionic transformations
\begin{align}
\theta' &= \theta + \eta(\tau + \theta \bar \theta) \cr
\bar \theta' &= \bar \theta \cr
\tau' &= \tau +  \bar \theta \eta 
\end{align}
and 
\begin{align}
\theta' &= \theta \cr
\bar \theta' &= \bar \theta + \bar \eta(\tau - \theta \bar \theta) \cr
\tau' &= \tau +  \theta  \bar \eta 
\end{align}

We can obtain the most general transformation by applying a fermionic transformation followed by a bosonic transformation.
We will come back to that later. 

The $\mathcal{N}=2$ transformations are a symmetry of the (\ref{EOMN2}) equations of motion (without the first derivative term)
with 
\begin{equation}
\mathcal{G}(\tau_1,\theta_1, \bar \theta_1;\tau_2,\theta_2, \bar \theta_2) = (D_{\bar \theta_1} \bar \theta'_1)^{\frac13} (D_{\theta_2} \theta'_2)^{\frac13} \mathcal{G}(\tau'_1,\theta'_1;\tau'_2,\theta'_2)
\end{equation}
Notice that the Jacobian factors are chiral and anti-chiral respectively. 

This follows from the observation that the chiral measure $d\tau d\theta\simeq d (\tau + \theta \bar \theta) d \theta$ transforms with a factor of $D_{\theta} \theta'$:
\begin{align}
\mathrm{Ber} \begin{pmatrix}\partial_\tau (\tau' + \theta' \bar \theta') &  \partial_\tau \theta'  \cr \partial_\theta (\tau' + \theta' \bar \theta') & \partial_\theta \theta'  \end{pmatrix} 
&= \mathrm{Ber} \begin{pmatrix}\partial_\tau (\tau' + \theta' \bar \theta')   & \partial_\tau \theta'  \cr D_\theta (\tau' + \theta' \bar \theta')   &  D_\theta \theta' \end{pmatrix} 
= (D_\theta\theta')^{-1} \mathrm{Ber} \begin{pmatrix}\partial_\tau (\tau' + \theta' \bar \theta')  & \partial_\tau \theta'  \cr 2 \bar \theta'  &  1\end{pmatrix}\cr 
&= (D\theta')^{-1}\left( \partial_\tau \tau'  - \partial_\tau \theta' \bar \theta' -\partial_\tau \bar \theta' \theta'    \right) = D_{\bar \theta} \bar \theta'
\end{align}

If we define the auxiliary (anti)chiral variables $\tau_{\pm} = \tau \pm \theta \bar \theta$, then the $\mathcal{N}=2$ super-reparameterizations
can be thought of as a subgroup of the product of groups of chiral and anti-chiral general coordinate transformations
\begin{align}
\tau_+ &\to \tau'_+(\tau_+, \theta) \qquad \qquad \theta \to \theta'(\tau_+, \theta) \cr
\tau_- &\to \tau'_-(\tau_-, \bar \theta) \qquad \qquad \bar \theta \to \bar \theta'(\tau_-, \bar\theta)
\end{align}
which satisfy $\tau' \equiv \tau'_+ - \theta' \bar \theta' = \tau'_- + \theta' \bar \theta'$.

The (\ref{EOMN2}) equations of motion (without the first derivative term) are actually invariant under the larger symmetry group, with
chiral and anti-chiral coordinate transformations acting separately on the two entries of the two-point function
\begin{align}\label{eq:extended}
\mathcal{G}(\tau_1,\theta_1, \bar \theta_1;\tau_2,\theta_2, \bar \theta_2) &= (D_{\bar \theta_1} \bar \theta'_1)^{\frac13} (D_{\theta_2} \theta''_2)^{\frac13} \mathcal{G}(\tau'_1,\theta'_1;\tau''_2,\theta''_2) \cr
\mathcal{\bar G}(\tau_1,\theta_1, \bar \theta_1;\tau_2,\theta_2, \bar \theta_2) &= (D_{\theta_1} \theta''_1)^{\frac13} (D_{\bar \theta_2} \bar \theta'_2)^{\frac13} \mathcal{\bar G}(\tau''_1,\theta''_1;\tau'_2,\theta'_2)
\end{align}
with $\tau'' = (\tau')^*$, etc. These extra transformations are incompatible with the UV boundary condition. 

Global super-conformal transformations are generated by super-translations, U(1) rotations and the inversion 
\begin{equation}
\tau \to \tau' =-\frac{1}{\tau}  \qquad \qquad \theta \to \theta' = \frac{\theta}{\tau} \qquad \bar \theta \to \bar \theta' = \frac{\bar \theta}{\tau} 
\end{equation}
Observe that the inversion maps $\tau_\pm \to -\frac{1}{\tau_\pm}$.
Obviously, super-conformal transformations only mix $\tau_+$ with $\theta$ and $\tau_-$ with $\bar \theta$. We can thus write
\begin{align}
\tau'_+ &= \frac{a \tau_+ + \alpha \theta + b}{c \tau_+ + \gamma \theta + d} \qquad \qquad \theta' = \frac{\beta \tau_+ + e \theta + \delta}{c \tau_+ + \gamma \theta + d} \cr
\tau'_- &= \frac{\bar a \tau_- + \bar \alpha \bar \theta + \bar b}{\bar c \tau_- + \bar \gamma \bar \theta + \bar d} \qquad \qquad \bar \theta' = \frac{\bar \beta \tau_- + \bar e \bar \theta + \bar \delta}{\bar c \tau_- + \bar \gamma \bar \theta + \bar d} 
\end{align}
These are sensible if and only if $\tau'_+ - \tau'_- = 2 \theta' \bar \theta'$, i.e. 
\begin{equation}
(a \tau_+ + \alpha \theta + b)(\bar c \tau_- + \bar \gamma \bar \theta + \bar d) - (\bar a \tau_- + \bar \alpha \bar \theta + \bar b)(c \tau_+ + \gamma \theta + d) = 2 (\beta \tau_+ + e \theta + \delta)(\bar \beta \tau_- + \bar e \bar \theta + \bar \delta)
\end{equation}
i.e.
\begin{align}
a \bar c - \bar a c &= 2 \beta \bar \beta \qquad 
b \bar d - \bar b d = 2\delta \bar \delta \qquad
\alpha \bar c - \bar a \gamma = 2 e \bar \beta \cr
a \bar \gamma - c \bar \alpha &= 2 \beta \bar e \qquad 
\alpha \bar d - \bar b \gamma = 2 e \bar \delta \qquad
b \bar \gamma - \bar \alpha d = 2 \delta \bar e \cr
a \bar d + b \bar c - \bar a d - \bar b c &= 2 \beta \bar \delta + 2 \delta \bar \beta \qquad 
a \bar d - \alpha \bar \gamma - b \bar c + \bar a d - \bar \alpha \gamma - \bar b c = 2 \beta \bar \delta + 2 e \bar e- 2 \delta \bar \beta
 \end{align}
i.e. in matrix form
\begin{equation}
\begin{pmatrix} a & c & \beta \cr b & d & \delta \cr \alpha & \gamma& e \end{pmatrix}\begin{pmatrix} 0 & 1 & 0 \cr -1 & 0 & 0 \cr 0 & 0& 2 \end{pmatrix}\begin{pmatrix}  \bar a &  \bar b &  \bar \alpha \cr \bar c & \bar d & \bar \gamma \cr  - \bar \beta & - \bar \delta & - \bar e \end{pmatrix} = \begin{pmatrix} 0 & x & 0 \cr -x & 0 & 0 \cr 0 & 0& 2x \end{pmatrix}
\end{equation}
where $x$ is undetermined and can be set to $1$ as a choice of overall normalization of the coefficients. 
They form an SU$(1,1|1)$ group with four bosonic generators and four fermionic generators.

The $\mathcal{N}=2$ super-Schwarzian derivative is 
\begin{equation}
S(\tau', \theta', \bar \theta';\tau, \theta, \bar \theta) = \frac{\partial_\tau \bar D \bar \theta'}{\bar D \bar \theta'} 
- \frac{\partial_\tau D \theta'}{D \theta'} - 2 \frac{\partial_\tau \theta' \partial_\tau \bar \theta' }{(\bar D \bar \theta')(D \theta')} = \cdots + \theta \bar \theta S_b(\tau', \theta', \bar \theta';\tau, \theta, \bar \theta)
\end{equation}
which satisfies a chain rule of the form 
\begin{equation}
S[\tau'', \theta'',\bar \theta'';\tau, \theta, \bar \theta] = \left(D \theta' \right) \left(\bar D \bar \theta' \right)S[\tau'', \theta'', \bar \theta'';\tau', \theta', \bar \theta'] + S[\tau', \theta', \bar \theta';\tau, \theta, \bar \theta] 
\end{equation}
The super-space action 
\begin{equation}
\int d \tau d \theta d \bar \theta S[\tau', \theta', \bar \theta';\tau, \theta, \bar \theta] = \int d\tau S_b(\tau', \theta', \bar \theta';\tau, \theta, \bar \theta)
\end{equation}
is a natural $\mathcal{N}=2$ supersymmetrization of the Schwarzian action. 

If we parameterize the super-Jacobians as 
\begin{equation}
D_\theta \theta' = \rho(\tau - \theta \bar \theta) (1 + \bar \theta \lambda) \qquad D_{\bar \theta} \bar \theta' = \bar \rho(\tau + \theta \bar \theta) (1 + \theta \bar \lambda)
\end{equation}
so that 
\begin{align}
\frac{\partial_\tau \theta'}{D_\theta \theta'} &= \theta \partial_\tau \log \rho(\tau) + \frac12 \frac{ \lambda(\tau +\theta \bar \theta)}{1 + \bar \theta \lambda} \cr
\frac{\partial_\tau \bar \theta'}{D_{\bar \theta} \bar \theta'} &=\bar \theta \partial_\tau \log \bar \rho(\tau) + \frac12 \frac{ \bar \lambda(\tau -\theta \bar \theta)}{1 + \theta \bar \lambda}
\end{align}
then we have 
\begin{equation}
S_b(\tau', \theta', \bar \theta';\tau, \theta, \bar \theta) =  \partial^2_\tau \log (\rho \bar \rho)
- \frac12( \partial_\tau \log (\rho\bar \rho))^2 +\frac12( \partial_\tau \log (\rho/\bar \rho))^2   -\frac12 \partial \lambda(\tau) \bar \lambda(\tau) + \frac12  \lambda(\tau) \partial \bar \lambda(\tau)\end{equation}

In order to go further, we need to pick a specific parameterization of the general super-reparameterization symmetry transformations. 
If we choose
\begin{align}
\theta' &= \rho(\tau + \theta \bar \theta) \left(\theta + \eta(\tau + \theta \bar \theta) \right) \cr
\bar \theta' &=\bar \rho(\tau - \theta \bar \theta) \left(\bar \theta + \bar \eta(\tau - \theta \bar \theta) \right) \cr
\tau' &= f(\tau) +  \theta \bar g(\tau) + \bar \theta g(\tau) + h(\tau) \theta \bar \theta
\end{align}
then we have 
\begin{align}
\bar g(\tau- \theta \bar \theta) + ( h(\tau)+ \partial_\tau f)  \bar \theta &= \bar \rho(\tau - \theta \bar \theta) \left(\bar \theta + \bar \eta(\tau - \theta \bar \theta) \right)  \left(  \rho(\tau- \theta \bar \theta)+ 2 \bar \theta \partial_\tau( \rho(\tau) \eta(\tau) \right) \cr
g(\tau+ \theta \bar \theta) + (- h(\tau)+ \partial_\tau f) \theta &= \rho(\tau +\theta \bar \theta) \left(\theta + \eta(\tau + \theta \bar \theta) \right)  \left(  \bar \rho(\tau+ \theta \bar \theta)+ 2  \theta \partial_\tau( \bar \rho(\tau) \bar \eta(\tau) \right)
\end{align}
i.e.
\begin{align}
g(\tau) &=  \rho(\tau) \bar \rho(\tau)\eta(\tau) \cr
\bar g(\tau) &=  \rho(\tau) \bar \rho(\tau)\bar \eta(\tau) \cr
\partial_\tau f &=\rho(\tau) \bar \rho(\tau) - \bar \rho(\tau) \bar \eta(\tau) \partial_\tau( \rho(\tau) \eta(\tau) ) -\rho(\tau) \eta(\tau) \partial_\tau( \bar \rho(\tau) \bar \eta(\tau) )\cr  
h(\tau)&=\rho(\tau) \eta(\tau) \partial_\tau( \bar \rho(\tau) \bar \eta(\tau) ) - \bar \rho(\tau) \bar \eta(\tau) \partial_\tau( \rho(\tau) \eta(\tau) ) 
\end{align}

The phase $\rho/\bar \rho = e^{2 i \sigma(\tau)}$ controls $U(1)_R$ rotations and defines an axion field. The norm 
$\rho \bar \rho$ equals $\partial_\tau f$ plus fermionic corrections: 
\begin{equation}
\partial_\tau f =\rho(\tau) \bar \rho(\tau)\left(1-\eta (\partial_\tau - i \partial_\tau \sigma) \bar \eta - \bar \eta (\partial_\tau+ i \partial_\tau \sigma) \eta\right)  
\end{equation}
Finally, $\lambda = 2 \frac{\partial_\tau( \rho \eta)}{\rho}$. 

Thus the bosonic part of the action consists of the usual Schwarzian plus a standard kinetic term for $\sigma$, with a specific 
relative coefficient:
\begin{equation}
S_b(\tau', \theta', \bar \theta';\tau, \theta, \bar \theta) = \frac{\partial^3 f}{\partial f} - \frac32 \left(\frac{\partial^2 f}{\partial f}\right)^2
-2( \partial_\tau \sigma)^2 + \cdots \end{equation}

 The relation between these two coefficients has some implications for the low energy near
extremal thermodynamics. 
    Setting  $f = \tan { \tau \over 2 } $, and setting $\tau = 2 \pi u /\beta $, where $u $ is physical euclidean
time, we get 
    \be \label{TDTwo}
    { \log Z \over N }  = { \alpha_s \over J } \int du   \left[ \{ f, u \} - 2 (\partial_u\sigma - i \mu)^2 \right] 
\to  { \alpha_s \over  J } \left[ {   2 \pi ^2 \over \beta}   + 2  \beta \mu ^2 \right]
\ee
where we also included a small chemical potential $\mu$ for the R-charge and we set $\partial_u \sigma =0$.  Small $\mu $ means that 
$\mu \ll J$, and we have $\beta \mu$ that can be of order one.  
     From this we can compute the energy and charge and entropy,  $\log Z = S - \beta E + \beta \mu Q_R $,  
    \be
     { E \over N } = \alpha_s \left(  { 2 \pi^2 \over \beta^2 J } + { 2  \over J } \mu^2  \right) ~,~~~~~~
     { Q_R \over N } = 4 { \alpha_s  \over J }  \mu ~,~~~~~~{ S \over N } = \alpha_s 4 \pi^2 { 1 \over \beta J } 
     \ee
    and we can express the entropy as a function of the energy and the charge as 
    \be \label{EntrNTwo}
    { S-S_0 \over N } =   \pi   \sqrt{ { 8 \alpha_s E \over J N } - {  } \left( { Q_R \over N } \right)^2 } 
    \ee
    where $S_0$ is the ground state entropy. 
This is correct only for very small values of the energy and the charge $ { E \over J N } \ll 1$ and  $Q_R/N \ll 1$ . 
    Recall that we are normalizing the charge of the fermion to $Q_R = \pm { 1 \over \hat q } $. 
This means that the period of the field $\sigma$ is  $\sigma = \sigma + 2 \pi \hat q $.  

   In any charged SYK model (expanded around a zero charge background) 
 we have  similar formulas but with an extra coefficient in front of the 
$(\partial_u \sigma - i \mu)^2 $ term. 
 ${\cal N}=2$ supersymmetry  fixes this extra coefficient.   
    
As in the discussion around \nref{ZeroMOne}, we can now consider the effects of the bosonic and 
fermionic zero modes. Since there is an equal number of boson and fermion zero modes in this case (four of
each) we find that there are no $\beta$ dependent prefactors in the low temperature  partition function 
($ 1\ll \beta J \ll N $)
\be \label{PartTwo}
 Z_{\rm 1-loop} \sim {  } e^{ S_0} e^{ N { \alpha_s \over  J } \left[ {   2 \pi ^2 \over \beta}   + 2  \beta \mu^2 \right] } 
\ee
This leads to the following prefactor in the density of states 
\be
D(E,Q_R) = \int d \beta  d\mu \beta e^{ \beta E - \beta \mu Q_R} Z(\beta , \mu ) \propto  { 1 \over (\Delta S)^2 } e^{S_0 + \Delta S }
\ee 
where $\Delta S = S - S_0 $ is given by the left hand side of \nref{EntrNTwo}.
 
In this case we do not expect the result \nref{PartTwo} to be exact. In fact, already we expect to be 
multiplied by a sum over ``windings'' of  the $\sigma$ rotor degree of freedom of the form 
\be
 \sum_{n=-\infty}^\infty   e^{ - { 2 \alpha_s \beta \over J } (  { 2 \pi \over \beta } \hat q n - i \mu)^2 }
\ee

\section{Four point function and the spectrum of operators } 
\label{4pt} 

\begin{figure}[h]
\center
\includegraphics[width=5.5in]{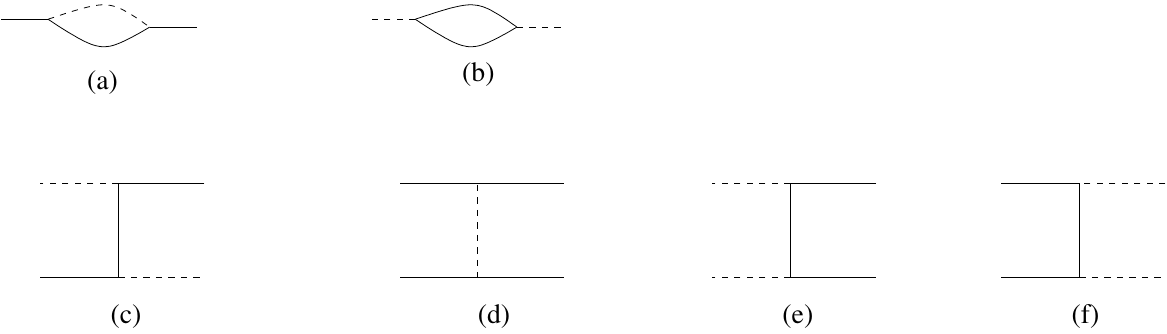}
\caption{ (a) Diagram contributing to a correction to the fermion propagator. Full lines are fermions and dotted lines are bosons. (b) Correction to the boson propagator. 
(c) A simple ladder diagram contributing to the four point function in the fermionic channel, where the intermediate state obtained  when we cut the ladder is a fermion. 
(d), (e), (f) Diagrams contributing in the bosonic channel, with either a pair of bosons or a pair of fermions. The full ladders are obtained by iterating these diagrams.
These are the diagrams for $\hat q=3$ and they look similar in the general case.  }
\label{Diagrams}
\end{figure}
The four point function can be computed by techniques similar to those discussed in \cite{kitaev2015talk,JPRV16,JMDS16}. 
We should sum a series of ladder diagrams, see figure \ref{Diagrams}. 
There are various types of four point functions we could consider. The simplest kind has the form 
\be
\langle \psi^i(\tau_1) \phi^i(\tau_2) \psi^j(\tau_3) \phi^j(\tau_4) \rangle 
\ee
In this case the object propagating along the ladder is fermionic, produced by a   boson and fermion operator.  
We will not present the full  form of the four point function in   detail, but we will note the dimensions
 of the operators appearing in the singlet channel OPE (the $\tau_1 \to \tau_2$ limit). 
As in  \cite{kitaev2015talk,JPRV16,JMDS16} these dimensions are computed by using conformal symmetry to diagonalize the ladder 
kernel in terms of a basis of functions of two variables 
with definite conformal casimir specified by a conformal dimension $h$. Then setting the kernel equal to one gives us the spectrum of dimensions that 
can appear in the OPE. The problem can be sepated into contributions where the intermediate functions are essentially  symmetric or antisymmetric under
the exchange of variables. 
This gives us two sets of fermionic operators specified by the conditions 
\bea  \la{Ksf}
1 & = & k_{s}(h)\equiv - 2^{ -1+ { 2 \over \hat q } } { \Gamma( 2 - { 1 \over \hat q } ) \over \Gamma(1 + { 1 \over \hat q } ) }
{ \Gamma( { 1 \over 4 } + { 1 \over 2 \hat q } - { h \over 2 } ) \Gamma( {1\over 4 } + { 1 \over 2 \hat q } + { h \over 2 } ) \over
\Gamma( { 5 \over 4 } - { 1 \over 2 \hat q } - { h \over 2 } ) \Gamma( { 1 \over 4 } - { 1 \over 2 \hat q } + { h \over 2 } )  }
\cr \la{Kaf}
1 & = & k_{a}(h)\equiv - 2^{ -1+ { 2 \over \hat q } } { \Gamma( 2 - { 1 \over \hat q } ) \over \Gamma(1 + { 1 \over \hat q } ) }
{ \Gamma( { 3 \over 4 } + { 1 \over 2 \hat q } - { h \over 2 } ) \Gamma( -{1\over 4 } + { 1 \over 2 \hat q } + { h \over 2 } ) \over
\Gamma( { 3 \over 4 } - { 1 \over 2 \hat q } - { h \over 2 } ) \Gamma( { 3 \over 4 } - { 1 \over 2 \hat q } + { h \over 2 } )  }
\eea
From the first and second we get eigenvalues of the form  
\bea
h_{s,m}& =&  { 3 \over 2} ,~3.3211..., ~5.2409..., ~,\cdots  ~~~~~ h_{s,m }= \Delta_\psi + \Delta_b + 2 m + \gamma_m   
\cr
h_{a,m} & =& { 3 \over 2 } ,~3.5659...,~5.5949...,~\cdots~~~~~~h_{a,m} = \Delta_\psi + \Delta_b + 2 m +1 - \tilde \gamma_m 
\eea
Except for $h=3/2$ the numbers do not appear to be rational. 
They approach the values we indicated above for large $m$, with small positive 
$\gamma_m$ or $\tilde \gamma_m$
for large $m$. These operators can be viewed as having the rough form $\psi^i \partial^{ n} \phi^i$ with $n=2m, ~2m+1$ respectively. 

It is also possible to look at the ladder diagrams corresponding to four point functions of the form $\langle \psi^i \psi^i \psi^j \psi^j\rangle$. When we compute
the ladders these are mixed with ones with structures like $\langle \psi^i \psi^i b^j b^j \rangle$ or $\langle b^i b^i b^j b^j \rangle$, 
 see figure \ref{Diagrams}.
 So the kernel even for a given intermediate $h$ is a two by two matrix. 
Diagonalizing this matrix we find that the operators split into two towers which are the partners of the above one. 
These bosonic partners   have conformal dimensions given by  $h_{s,m} + \half$ and $h_{a,m} - \half $ for each of the two fermionic towers. 
  Of course, it should be possible to directly use super-graphs so that we can preserve manifest supersymmetry. 
 
 Now, we expect that the case where $h_{s}=3/2$ and its bosonic partner with $h=2$   lead to a divergence in the computation of the naive expression 
 for the four point function and that the proper summation would reproduce what we obtain from the super-Schwarzian action discussed in section 
\ref{SuperSchwarzian}.

The pair of modes with $h_{a}=3/2$ and its bosonic partner at $h=1$ are more suprising. The origin of the $h=1$ mode is due
to the rescaling symmetry of the IR equations mentioned in \nref{symNai}.  In fact, one can extend that symmetry to a local symmetry of the form
\be \la{localRes}
G_{\psi \psi}(t,t') \to \lambda(t)\lambda(t') G_{\psi \psi}(t,t') ~,~~~~~~ G_{bb}(t,t') \to [ \lambda(t) \lambda(t')]^{ 1 -\hat q } G_{bb}(t,t')
\ee
 which would naively suggest the
presence of an extra set of zero modes. 
However, we noted that this symmetry is broken by the UV boundary conditions. Of course this was also true of the reparameterization symmetry.
However, \nref{localRes}    changes the short distance form of the correlators, which leads us to expect terms in the effective action of the form 
$J \int d\tau (\lambda(t) - 1)^2$, which strongly suppress the deviations from the value of $\lambda$ given by the   short distance solution.  
Thus, in the low energy theory we do not expect a zero mode from these.
Indeed, when we look at the ladders with the boson exchanges, we see  that the basis of functions we are summing over when we express the four point function
should be the same as the one for the usual SYK model, (see \cite{JMDS16}). Namely, the expansion for the four point function can be expressed as an integral 
over $h=1/2 + i s$ and a sum over even values of $h$.   Since $h=1$ is not even, it does not lead
to a divergence. Then we conclude that it corresponds to an operator of the theory. 
  It  looks like a marginal deformation, since it has  $h=1$.  In the UV, it looks like the operator corresponds to a relative rescaling of the boson and fermion field. 
  We think that the transformation simply corresponds to a rescaling of $J$, which breaks the original supersymmetry but preserves a new rescaled supersymmetry. 
  We have not studied in detail the meaning of its supersymmetric partner which is a dimension 3/2 operator. 
  

The case with ${\cal N}=2$ supersymmetry leads to similar operators in the singlet channel with zero $U(1)_R$ charge. 
The fermions have the same dimensions as in \nref{Ksf}, \nref{Kaf}, but each with a factor of two degeneracy arsing from the fact that now we 
change $\psi^i b^i \to \psi^i \bar b^i$ and $\bar \psi^i b^i $. The  the bosonic operators fill a whole ${\cal N}=2$ multiplet with dimensions
$(h_{s,m} - \half, h_{s,m}, h_{s,m} + \half)$ and $(h_{a,m} - \half, h_{a,m}, h_{a,m} + \half) $. In this model the functions we need to sum over in order to get the
four point function are more general than the ones in the SYK model, since now that the basic two point function $\langle \psi^i(t_1) \bar \psi^i(t_2) \rangle$ 
does not have a definite symmetry. So now the expression for the four point function should include a sum over all values of $h$, includding both even and odd values, 
depending on whether we consider symmetric or antisymmetric parts. Though we have not filled out all the details we expect that by supersymmetry we will have
that the multiplet coming from the symmetric tower with dimensions $(1,3/2,2)$ should lead to the superschwarzian while the second one, coming from $h_{a,m}$, 
also with dimensions $(1,3/2,2)$ should correspond to operators in the IR theory. As before these arise from symmetries of the low energy equations, namely (\ref{eq:extended}). 
Let us discuss in detail the ones corresponding to the dimension two operators. 
The low energy equations have the form 
\be \label{GN2IR}
G_{b \bar b} *   G_{\bar \psi \psi }^{\hat q -1} = -\delta  ~,~~~~~~~~ G_{\psi \bar \psi} * [ (\hat q-1  ) G_{\bar b , b } G_{\bar \psi \psi }^{\hat q -2} ] = -\delta 
\ee
where $*$ is a convolution and we think of each side as a function of two variables. The right hand side is a delta function that sets these two variables equal.  
We also have   complex conjugate equations obtained by replacing $G_{\psi \bar \psi} \leftrightarrow G_{\bar \psi \psi } $, $G_{b \bar b} \leftrightarrow G_{\bar b b } $. 
We can then check that the following is a symmetry 
\bea
G_{\psi \bar \psi } & \to  G'_{\psi \bar \psi } (\tau_1 , \tau_2) = [f'(\tau_1) h'(\tau_2)]^{\Delta_\psi } G_{\psi \bar \psi } (f(\tau_1) , h(\tau_2))
\cr
G_{\bar \psi   \psi } & \to  G'_{\bar \psi   \psi } (\tau_1 , \tau_2) = [h'(\tau_1) f'(\tau_2)]^{\Delta_\psi } G_{\bar \psi   \psi } (h(\tau_1) , f(\tau_2))
\cr
G_{b \bar b } & \to  G'_{b\bar b } (\tau_1 , \tau_2) = [f'(\tau_1) h'(\tau_2)]^{\Delta_b } G_{b\bar b } (f(\tau_1) , h(\tau_2))
\eea
and similarly for $G_{\bar b b}$. If $G$ is a solution of \nref{GN2IR}, then $G'$ is also a solution.  
The  reparameterizations which are nearly zero modes of the full problem are those that obey $h = f$. The ones where they are different, are far from being zero modes of 
the full problem. The reality condition sets  that $h(\tau) = f(\tau)^*$. These look similar to two independent coordinate transformations that preserve 
conformal gauge in a two dimesional space,  with a  boundary condition that restricts them to be equal.

\section{Conclusions}

We have studied  supersymmetric generalizations of the SYK model. We studied models with ${\cal N} =1,2$ supersymmetry. 
Both  models are very similar to the SYK system, with a large ground state entropy and 
 a   large $N$ solution that is scale invariant in the IR. In these super versions, the scale invariance becomes a superconformal symmetry and the leading order 
 classical solutions preserve supersymmetry. These large $N$ solutions were also checked against numerical exact diagonalization results. 
 As in SYK, there is also 
 an emergent superconformal symmetry that is both spontaneously and explicitly broken. This action gives the leading corrections to the low energy thermodynamics and
 should produce the largest contributions to the four point function. Besides the ordinary reparameterizations, we have fermionic degrees of freedom and, in the 
 ${\cal N}=2$ case, a bosonic degree of freedom associated to a local $U(1)$ symmetry, which is related to the $U(1)_R$ symmetry. 
  A similar bosonic degree of freedom arises in other situations with a $U(1)$ symmetry, such as the model studied in \cite{SS15}. Here supersymmetry implies 
  that the coupling in front the schwarzian action is the same as the one appearing in front of the action for this other bosonic degree of freedom. This fixes the low energy thermodynamics in terms of only one 
overall coefficient, see \nref{TDTwo}. 
   
  We also analyzed the operators in the ``singlet'' channel. These operators have anomalous dimensions of order one. Therefore, in these models, supersymmetry is
  not enough to make those dimensions very high. 
  
 In the ${\cal N}=1$ case, the exact diagonalization results allowed us to show that the ground state energy is non-zero and of order $E_0 \propto e^{ - 2 S_0 }$. 
 This means that supersymmetry is non-perturbatively broken. 
 On the other hand,  in the ${\cal N}=2$ case, supersymmetry is {\it not} broken and there is a large number of zero energy states which matches the ground state
 entropy computed using the  large $N$ solution. 
  Furthermore, these zero energy states can have non-zero R charge, but with an R charge parametrically smaller than $N$, and even smaller than one. 
 
 These results offer some lessons for the study of supersymmetric black holes. In supergravity theories there is a large variety of extremal black holes that are supersymmetric in the gravity approximation. The fact that supersymmetry can be non-perturbatively broken offers a cautionary tale for attempts to reproduce 
 the entropy using exactly zero energy states (see eg. \cite{Kinney:2005ej,Chang:2013fba}). Of course, in situations where there is an index reproducing the entropy, as in \cite{Strominger:1996sh},
  this is not 
 an issue.  
 The authors of \cite{Benini:2015eyy}  have argued that the ground states of supersymmetric black holes carry zero R charge, where the $R$ charge is the IR one 
 that appears in the right hand side of the superconformal algebra. In our case there is only one continuous $U(1)_R$ symmetry and we find that
 the ground states {\it do not } have exactly zero charge. 
  A possible loophole is that the R-symmetry appearing in the superconformal algebra leaves invariant the thermofield double, not each copy individually. 
 Perhaps a modified version of the argument might be true since in our case the R charges are relatively small. Also the discrete chemical potential we 
 introduced in \nref{WittIn}, looks like a discrete version of the maximization procedure discussed in \cite{Benini:2015eyy}.  It seems that this is a point that could be 
 understood further. 
 
 Another surprise in the model is the emergence of additional local symmetries of the equations, beyond the ones associated to super-reparameterizations. 
 Similar symmetries arise in some of the non-supersymmetric models discussed in \cite{Gross:2016kjj}. 
 A common feature of these IR symmetries is that they change the short distance structure of the bilocals. Namely, they change the functions $G(t,t')$ even when 
 $t \to t'$. Since this is a region where the conformal approximation to the effective action develops divergencies, we see that now these divergencies will depend 
 on the symmetry generator. For this reason these symmetries do not give rise to zero modes, but are related to operators of the IR theory. 
 Amusingly, in the ${\cal N}=2$ case we also have an additional reparameterization symmetry of this kind. This symmetry together with the usual reparameterization
 symmetry look very similar to the conformal symmetries we would have in two dimensional $AdS_2$ space in conformal gauge. 
  
  We can wonder whether we can get models with ${\cal N} > 2$ supersymmetry.  It would be interesting to see if one can find models of this sort with only fermions. 
  A model with ${\cal N}=4$ supersymmetry that also involves dynamical bosons was studied in \cite{Anninos:2016szt}.

\subsection*{Acknowledgements}

We would like to thank D. Stanford, D. Simmons-Duffin, G. Turiaci and S. Yankielowicz for discussions. 

The research was supported by the NSF under Grant DMR-1360789 and by MURI grant W911NF-14-1-0003 from ARO.
Research at Perimeter Institute is supported by the Government of Canada through Industry Canada and by the Province of Ontario 
through the Ministry of Research and Innovation. SS also acknowledges support from Cenovus Energy at Perimeter Institute.
J.M. is supported in part by U.S. Department of Energy grant
de-sc0009988 and by the Simons Foundation grant 385600.

\bibliography{supersyk}

\end{document}